\documentclass[aps,prb,preprint,showpacs,floatfix]{revtex4-1}  % for PRB format
\usepackage{amsmath,amssymb,xcolor}
\usepackage[colorlinks,bookmarks=false,citecolor=red,linkcolor=blue,urlcolor=blue]{hyperref}
\usepackage{graphicx,epstopdf}
\usepackage{hyperref}
\hypersetup{
    colorlinks,
    citecolor=blue,
    filecolor=blue,
    linkcolor=blue,
    urlcolor=blue}
\begin{document}
\title{Correlated spin liquids in the quantum kagome antiferromagnet at finite field: a renormalisation group analysis} 
\author{Santanu Pal}
\email{sp13rs010@iiserkol.ac.in}
\author{Anirban Mukherjee}
\email{am14rs016@iiserkol.ac.in}
\author{Siddhartha Lal}
\email{slal@iiserkol.ac.in}
\affiliation{Department of Physical Sciences, Indian Institute of Science Education and Research-Kolkata, W.B. 741246, India}
%===============================================================
\begin{abstract}
We analyse the antiferromagnetic spin-$1/2$ XXZ model on the kagome lattice at finite external magnetic field with the help of a nonperturbative zero-temperature renormalization group (RG) technique. The exact nature of the ground and excited state properties (e.g., gapped or gapless spectrum etc.) of this system are still debated. Approximate methods have typically been adopted towards understanding the low-energy spectrum. Following the work of Kumar \emph{et al} (Phys. Rev. B {\bf 90}, 174409 (2014)), we use a Jordan-Wigner transformation to map the spin problem into one of spinless fermions (spinons) in the presence of a statistical gauge field, and with nearest-neighbour interactions. While the work of Kumar \emph{et al} was confined mostly to the plateau at $1/3$-filling (magnetisation per site) in the XY regime, we analyse the role of inter-spinon interactions in shaping the phases around this plateau in the entire XXZ model. The RG phase diagram obtained contains three spin liquid phases whose position is determined as a function of the exchange anisotropy and the energy scale for fluctuations arising from spinon scattering. Two of these spins liquids are topologically ordered states of matter with gapped, degenerate states on the torus. The gap for one of these phases corresponds to the one-spinon band gap of the Azbel-Hofstadter spectrum for the XY part of the Hamiltonian, while the other arises from two-spinon interactions. The Heisenberg point of this problem is found to lie within the interaction gapped spin liquid phase, in broad agreement with a recent experimental finding. The third phase is an algebraic spin liquid with a gapless Dirac spectrum for spinon excitations, and possess properties that show departures from the Fermi liquid paradigm. The three phase boundaries correspond to critical theories, and meet at a $SU(2)$-symmetric multicritical point. This special critical point agrees well with the gap-closing transition point predicted by Kumar \emph{et al}. We discuss the relevance of our findings to various recent experiments, as well as results obtained from other theoretical analyses. 
\end{abstract}
\pacs{}
\maketitle
%===========================================================================================
\section{Introduction}
Geometrically frustrated spin systems have drawn considerable attention ever since Anderson's seminal work highlighted the connection between the resonating valence bond (RVB) ground states and high temperature superconductivity~\cite{anderson1987resonating}. These systems are known to show exotic ground and excited state properties, such as liquid like ground states, fractional excitations etc.~\cite{balents2010spin,lee2008end}. The nearest-neighbour (n.n.) $S=1/2$ kagome antiferromagnet (KA) is one such promising model system in the search for unconventional states of matter. Despite sustained interest, the precise nature of the ground state and excitations of this system remain uncertain. Some analytical and numerical approaches have predicted a spin liquid ground state for the KA model Hamiltonian with a gapped excitation spectrum~ \cite{yan2011spin,waldtmann1998first}, while other studies support gapless excitations~\cite{PhysRevB.77.224413,PhysRevB.75.184406,PhysRevX.7.031020,sakai2017gapless,PhysRevB.97.104401}. Encouragingly, the material Herbertsmithite (ZnCu$_3$(OH)$_6$Cl$_2$) can be mapped to the n.n. $S=1/2$ KA model. Here too, however, conclusive results have remained elusive thus far: some experimental results on this material appear to show the existence of gapless excitations~\cite{PhysRevLett.100.157205,PhysRevLett.110.207208}, while some others suggest a gapped spin liquid ground state with fractional excitations~\cite{han2012fractionalized,fu2015evidence}. The materials Volborthite and Vesignieite~\cite{PhysRevLett.114.227202,PhysRevB.83.180407,PhysRevB.98.020404,PhysRevB.96.180413}, as well as the organic compound Cu-titmb~\cite{narumi2004observation}, are also believed to be described by the $S=1/2$ Heisenberg KA.  
\par
Further, in the presence of an external magnetic field, frustrated antiferromagnetic spin systems sometimes display the phenomenon of magnetization plateaux~\cite{honecker2004magnetization,PhysRevLett.102.075301,PhysRevB.83.180407,PhysRevB.92.174402}. For the $S=1/2$ KA, a plateau at a magnetisation per site of $1/3$ is well studied and predicted to be robust~\cite{nishimoto2013controlling,PhysRevB.71.144420,pal2018non}. Indeed, such a robust plateau at $1/3$ has been observed in the Volborthite material system~\cite{PhysRevLett.114.227202,PhysRevB.83.180407,
PhysRevB.98.020404,PhysRevB.96.180413}, as well as in Cu-titmb~\cite{narumi2004observation}. For a XXZ-KA model with XY anisotropy, a topological fluid state  with spin Hall conductivity $\sigma^{s}_{XY}=\frac{1}{4\pi}$ was proposed from a Chern-Simons Ginzburg-Landau mean field analysis together with a random phase approximation (RPA) analysis of fluctuations~\cite{PhysRevB.90.174409}. No firm conclusions could, however, be reached from this analysis for the regime of Ising anisotropy, with the physics of the Heisenberg point also remaining inaccessible.  Further studies of the influence of the Ising correlations is, therefore, required in order to reach a comprehensive understanding of the entire problem. This is a task we propose to undertake in this work.
\par
A strategy that can be adopted towards understanding the complex physics of the n.n. $S=1/2$ KA involves a mapping from the original spin problem to one of fermionic spinons via a Jordan-Wigner transformation~\cite{jordanWigner,PhysRevLett.63.322,PhysRevB.44.5246}. In two spatial dimensions, this transformation couples spinless fermions (i.e., the spinon excitations of the spin system) to a Chern-Simons statistical gauge field~\cite{PhysRevB.90.174409}, allowing for a mean-field treatment of the effects of the gauge field via the {\it average field approximation}.  Following Kumar \emph{et al.}~\cite{PhysRevB.90.174409}, for the spinon filling corresponding to a magnetisation per site of 1/3 filling, we begin by computing the dispersion spectrum for the reduced magnetic Brilloun zone (MBZ) comprising of nine bands. At this special filling, the effective chemical potential for the spinons is placed in the gap between the third and fourth bands (i.e., the first three bands are completely filled and the others empty). We then identify two points in the reduced MBZ as possessing the minimum energy gap between the top-most filled band (3rd band) and lowest empty band (4th band), thereby leading to the construction of an {\it effective two-patch problem}. We then analyse the fate of single spinon gap under two-spinon interactions via a nonperturbative renormalisation group (RG) analysis developed recently by some of us~\cite{mukherjee2018scaling}. 
\par 
By studying the XXZ Hamiltonian for the $S=1/2$ KA, the RG analysis opens new possibilities for the emergence of non-trivial liquid-like ground states at various values of the anisotopy parameter ($\lambda$ in equn.(\ref{eqn:Hamiltonian}) below) beyond those observed in an earlier study by Kumar \emph{et al.}~\cite{PhysRevB.90.174409}. Specifically, we have found three different phases for various ranges of the bare anisotropy $\lambda$. Of these, two are gapped topologically ordered spin-liquid phases, of which one corresponds to the one-spinon gapped state found in Ref.(\cite{PhysRevB.90.174409}), and the other possessing a gap arising purely from two-spinon scattering.  The non-trivial topological features of the two gapped phases can be distinguished using topological quantum numbers including a Chern number as well as the Volovik invariant~\cite{volovik2003universe,RevModPhys.82.1959}. By connecting the spin-flip part of the effective Hamiltonian (obtained at the stable fixed point of the RG calculation) with a boundary-condition changing twist operator~\cite{lieb1961two}, we are able to show that both gapped phases 
possess twofold ground-state degeneracy on the torus, and with fractional excitations interpolating between them. On the other hand, the third phase obtained from the RG corresponds to an algebraic spin-liquid state with gapless spinons possessing a Dirac-like dispersion. Here, spinon interactions reduce the effective Fermi velocity under renormalisation, effectively flattening the conical Dirac spectrum somewhat. The spinon self-energy, quasiparticle residue and lifetime from the RG indicate the existence of an unusual Fermi liquid in this phase. Finally, we also obtain the effective critical theories for the phase boundaries from the RG, revealing the existence of a non-trivial multicritical point lying at the intersection of the three spin liquid phases. In this way, the RG phase diagram obtained (see Fig.(\ref{fig:phasediagram})) represents a considerable advance in our understanding of the exotic phases of matter that are emergent in a prototypical geometrically frustrated spin system at finite magnetic field.
\par
This paper is organized as follows. In section-II, we transform the XXZ-spin Hamiltonian for the KA into that of fermionic spinons via 2D Jordan-Wigner transformation, and discuss the possible magnetization plateaux protected purely by a one-spinon gap. The computed dispersion spectrum in the reduced MBZ shows the gapped plateaux that are robust in the thermodynamic limit. Section-III will be devoted towards the construction of an effective pseudospin problem enabling the treatment of spinon interactions via a RG analysis. In section-IV, we present the RG flow equations for single-spinon gap and unveil the mechanism that leads to  the closing of this gap. We extend the RG analysis to also find the physics responsible for the opening of a two-spinon gap opening, as well as a phase that supports gapless spinons with a Dirac dispersion. Finally, by computing the effective Hamiltonians for the critical phase boundaries,   we reveal the entire phase diagram obtained from our RG analysis. In section-V, we define the different topological quantities that are employed in distinguishing the different phases. In section-VI, we calculate the self-energy, quasiparticle residue and lifetime for spinon excitations of the gapless phase. We conclude in Section-VII with some discussions. Finally, we present detailed calculations for various sections in the appendices. 
%================================================================================
\section{Fermionised Kagome XXZ model at finite field}
\begin{figure}[h]
\centering
\includegraphics[scale=0.35]{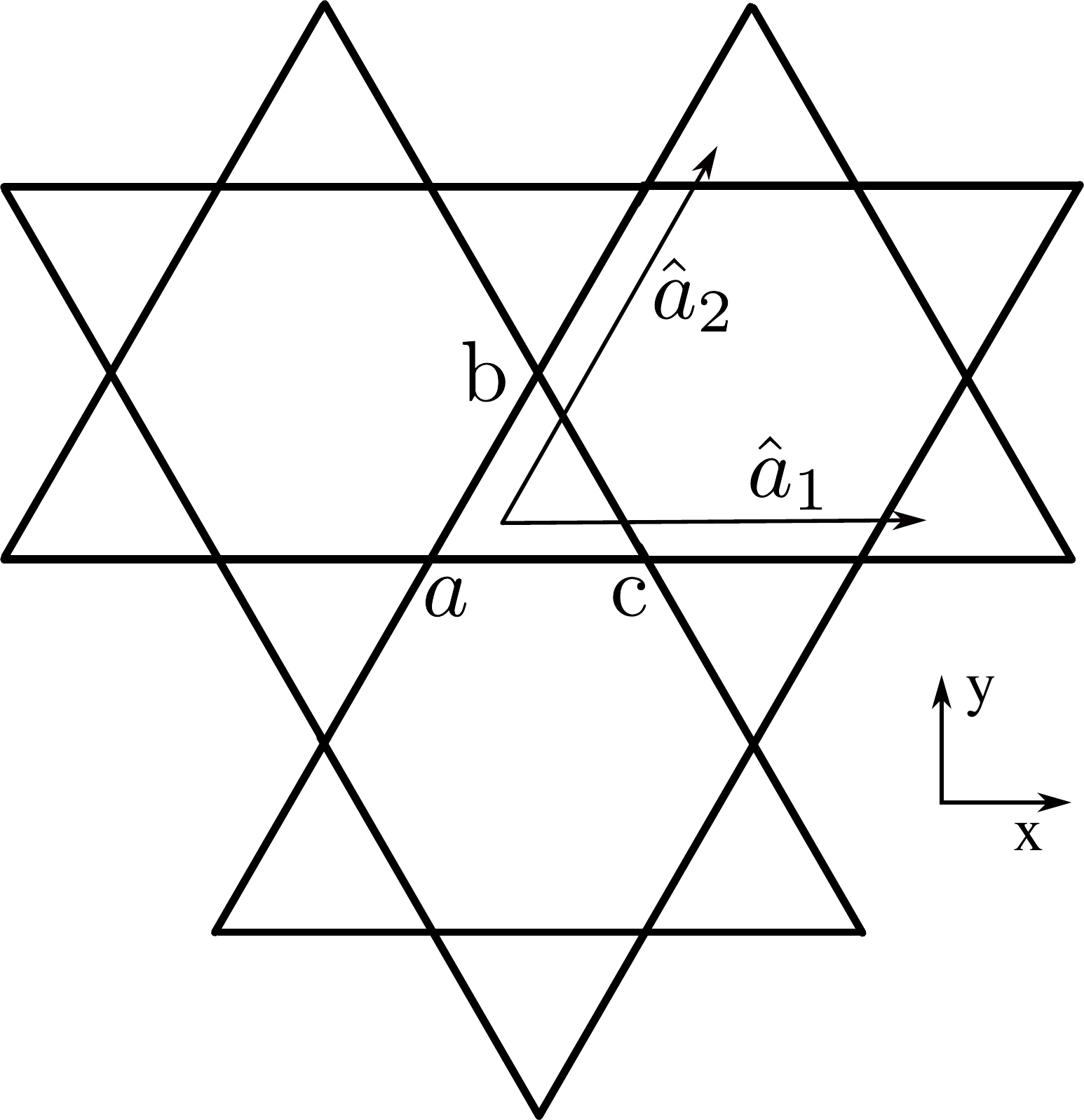}
\caption{The kagome lattice with basis vectors $\hat{a}_{1}=\hat{x}$ and $\hat{a}_{2}=\frac{1}{2}\hat{x}+\frac{\sqrt{3}}{2}\hat{y}$, such that the distance between nearest neighbour sites is fixed to half of the length of a basis vector, say, $|\hat{a}_{1}|$. Every triangular unit cell has three different sites marked by $a, b$ and $c$ indices.}
\label{fig:kagome}
\end{figure}
The $S=1/2$ antiferromagnetic  XXZ Hamiltonian for the kagome lattice in the presence of an external magnetic field ($h$) can be written as~\cite{PhysRevB.77.224413},
\begin{eqnarray}
H=J \sum_{<\vec{r}\vec{r}'>}[ S^x_{\vec{r}}S^x_{\vec{r}'}+S^y_{\vec{r}}S^y_{\vec{r}'}+\lambda S^z_{\vec{r}}S^z_{\vec{r}'}] - h\sum_{\vec{r}} S^z_{\vec{r}}~,
\label{eqn:Hamiltonian}
\end{eqnarray}
where $J$ is the exchange constant and $\lambda$ is the magnetic anisotropy between Ising and XY terms ($\lambda=1$ is the Heisenberg point). The sum is taken over nearest neighbour sites and $\vec{r}\in (\vec{R},i)$, where $\vec{R}=n_1\hat{a}_1+n_2\hat{a}_2$ ($n_1, n_2$ are integer) corresponds to the lattice vector for three sub-lattice  unit cell (up triangles). Further, $\hat{a}_1$ and $\hat{a}_{2}$ are the basis vectors and $i\in{(\text{a, b, c})}$ denote the three sub-lattices (see Fig.(\ref{fig:kagome})). The Hamiltonian is invariant under lattice translation along any basis vector of the lattice.
\par
Recently, we developed a non-perturbative approach based on twist operators towards obtaining firm criteria for the nature of the ground state of the kagome Heisenberg quantum antiferromagnet at zero as well as non-zero field~\cite{pal2018non}. In the present work, we aim to further develop a detailed understanding of the kagome problem at finite magnetic field via
a complementary non-perturbative renormalisation group (RG) analysis. In this way, we will carefully assess the role played by the Ising part of inter-spin interactions in determining the stability, as well as the properties, of the many-body ground state. In doing so, we rely on the results obtained by Fradkin and collaborators, who showed that Chern-Simons (CS) topological gauge-field theories can be obtained for gapped states corresponding to certain rational fractions of the magnetisation~\cite{PhysRevB.90.174409}. Thus, for a microscopic approach to the origin of different ground state properties at finite magnetisation, we begin by mapping the problem of interacting quantum spins onto a system of fermionic spinons coupled to a CS gauge field (with statistical angle $\theta =1/2\pi$) via a Jordan-Wigner (JW) transformation in two spatial dimensions~\cite{PhysRevLett.63.322,PhysRevB.90.174409}. For details of the JW transformation, we refer the reader to the detailed discussion in Kumar \emph{et al.}\cite{PhysRevB.90.174409}. 
\par
Upon performing the JW transformation, the fermionic Hamiltonian (\ref{eqn:Hamiltonian}) takes the form
\begin{eqnarray}
H &=&  \frac{J}{2}\sum_{<\vec{r}\vec{r}'>}[\psi^\ast(\vec{r},t)e^{iA(\vec{r},t)}\psi(\vec{r}',t)+\text{h.c.}] 
+ J\lambda \sum_{<\vec{r}\vec{r}'>} \big(\frac{1}{2}-n(\vec{r},t)\big)\big(\frac{1}{2}-n(\vec{r}',t)\big)~,
\label{eqn:fermionic}
\end{eqnarray}
where $n(\vec{r},t)=\psi^\ast(\vec{r},t)\psi(\vec{r},t)$ is the fermion density operator, $A(\vec{r},t)$ is the spatial part of the CS statistical gauge field employed in mapping spins to fermionic spinons.
\begin{figure}[h!]
\centering
\includegraphics[scale=.4]{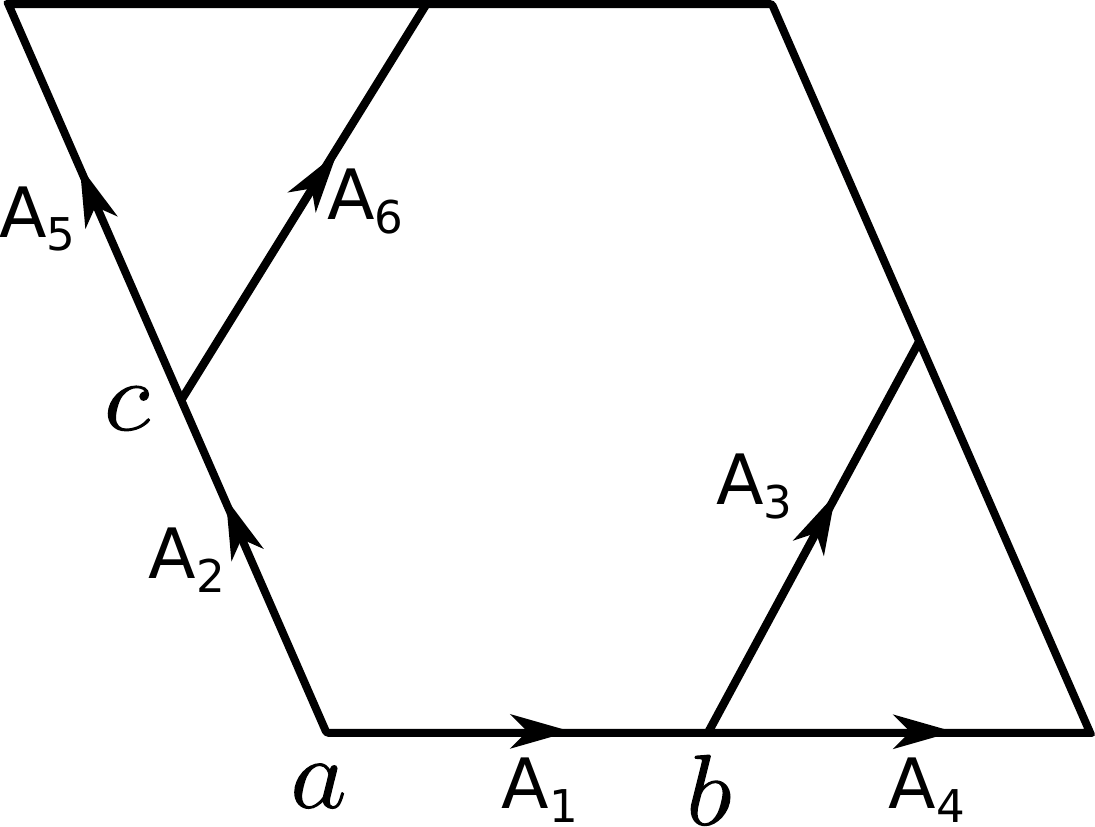}
\caption{Magnetic  unit cell for kagome lattice displaying the gauge field (A) along each link. The gauge choice A$_1=$A$_3=$A$_4=$0, A$_2=\phi$, A$_5=-\phi+3\phi x_1$ and A$_6=3\phi x_1$ ($x_1$ is the coordinate vector along $\hat{a}_1$ direction) gives a flux of $1/3$ unit flux quantum within every plaquette.}
\label{fig:magunitcell}
\end{figure}
We first obtain the dispersion spectra of the non-interacting spinons on the kagome lattice from the XY-part of the Hamiltonian by putting $\lambda=0$. It is important, however, to remember that the spinons remain coupled to the statistical gauge field in eqn.(\ref{eqn:fermionic}). A mean-field ansatz then involves considering a uniform flux in every plaquettes with the  gauge choice as shown in Fig.(\ref{fig:magunitcell}), and ignoring the fluctuations of the gauge field. One then obtains plateaux of the magnetization corresponding to 1/3, 2/3, 5/9 etc. Indeed, the plateaux are associated with a uniform flux of $\phi=2\pi \frac{p}{q}$ with $p, q\in \mathbb{Z}$, and correspond to an average filling of $<n>=\frac{p}{q}$ (where $q$ is the periodicity of magnetic unit cell). From the associated Hofstadter spectrum for the free spinon problem, the plateaux at 1/3 is found to be the most robust (i.e., it is stabilised by the largest single-spinon gap), and corresponds to 1/3-filling (average site occupancy) of each sublattice in the unit cell~\cite{PhysRevB.90.174409}. One can obtain the dispersion spectrum for the $<n>=\frac{1}{3}$ state by solving the associated Harper's equation numerically for the three sub-lattices~\cite{li2011tight,jiang2006mobility,PhysRevB.90.174409}. Hereafter, we assume a value of the exchange coupling $J=1$. As the periodicity of magnetic unit cell is $q=3$, a spinon band associated with a given sub-lattice is further split into three bands, giving a total of nine bands in the spinon spectrum. For a filling of $1/3$, the lowest three bands are filled and the rest empty, i.e., the effective chemical potential (equivalent to the magnetic field ($h$) of the original spin problem) is placed midway between the third and fourth spinon bands. Therefore, in analysing the effects of spinon scattering at low energy scales via a RG formalism, it is sufficient to focus on the 3rd (completely filled) and 4th (empty) bands. 
\begin{figure}[h!]
\centering
\includegraphics[scale=.65]{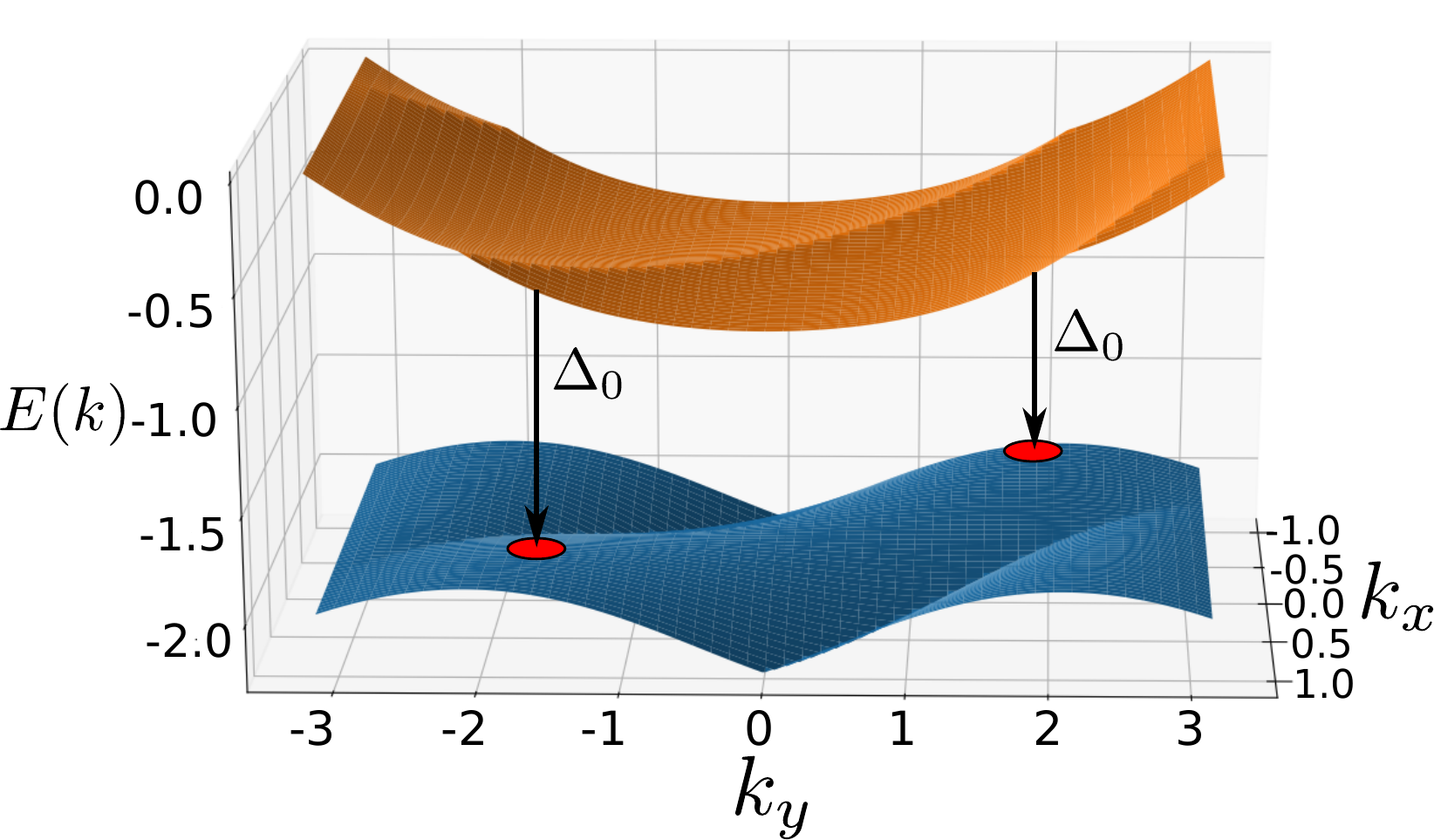}
\caption{Dispersion spectrum of the kagome lattice for $\phi=2\pi/3$, where the lower (blue) and upper (orange) bands represent the topmost filled and lowest unfilled bands of the magnetic unit cell respectively. Here, $k_x$, $k_y$ and $E(k)$ are momenta and energy indices. At 1/3 flux, the enlargement of the magnetic unit cell reduces the magnetic Brillouin zone reduces in the $k_x$ direction. The red circles locate the two minimum energy-difference points ($\vec{k}=(0.26, -1.88)$ and $\vec{k}=(-0.25, 1.92)$) between the two bands, corresponding to an energy gap $\Delta_0=1.355$ (in units of the exchange constant $J$).}
\label{fig:dispersion}
\end{figure} 
\par
Fig.(\ref{fig:dispersion}) shows the dispersion spectrum at $1/3$-filling in the 3rd and 4th bands, clearly indicating the one-spinon spectral gap of the free spinon problem. From a finite size scaling analysis, we find that the minimum gap between the third and fourth bands saturates at a value of $\Delta_{0}=1.355\pm0.001$ (in units of $J$) in the thermodynamic limit at the corresponding momentum coordinates of $(k_x,k_y)=(0.26155, -1.88479)$ and $(-0.25148, 1.91703)$ (see Fig.(\ref{fig:dispersion})). We recall that the effect of Ising interactions on the one-particle spectrum was analysed in Ref.(\cite{PhysRevB.90.174409}) at the level of mean-field theory together with the effects of saddle-point fluctuations of the gauge degrees of freedom. This analysis concluded that the gap would close at $\lambda^{*}=0.6$. It, however, was unable to reach any firm conclusions on the nature of Ising dominated phase lying at couplings $\lambda >\lambda^{*}$. In the following sections, we conduct a RG analysis of the quantum fluctuations arising from the interplay of the XY and Ising terms. This RG phase diagram obtained will clarify the nature of various gapless and gapped phases as well as the transitions between them.  
%====================================================================================
\section{Effective two-patch problem for $1/3$-plateau}\label{twopatch}
\begin{figure}[h!]
\centering
\includegraphics[scale=.35]{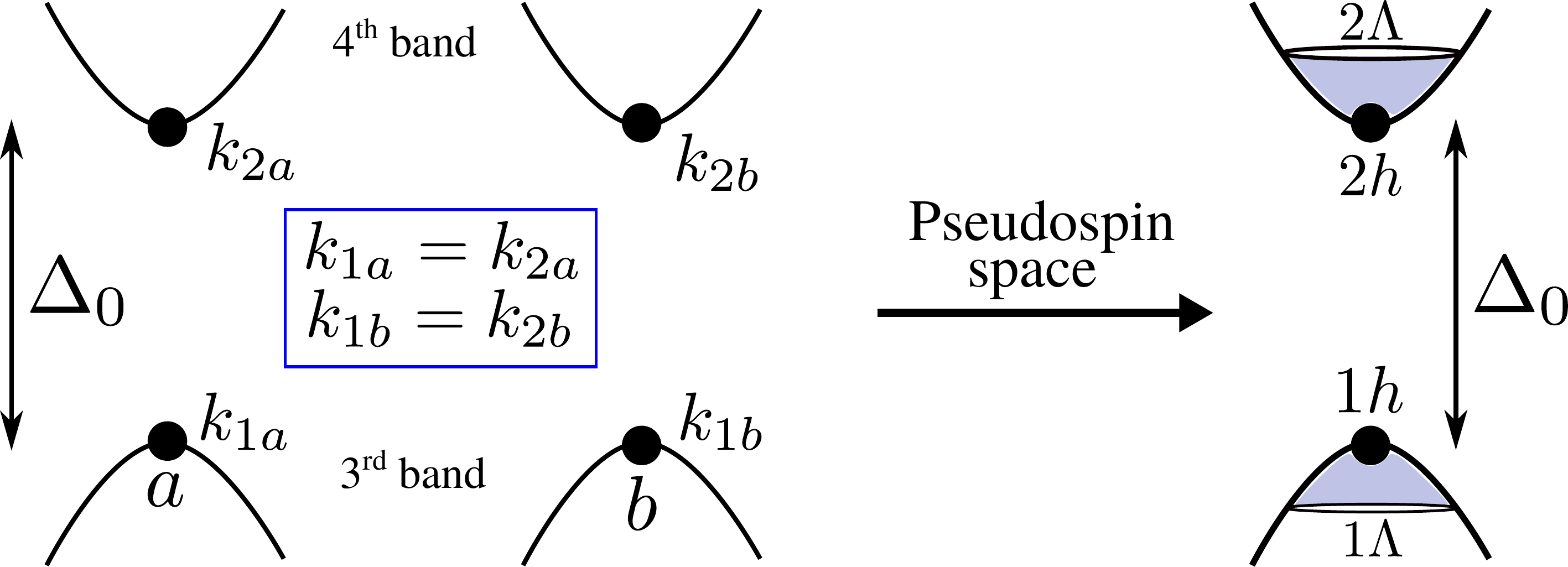}
\caption{Schematic digram of the two-patch problem, where the $a$ and $b$-points (black circles) corresponds to the two points with minimum energy gap $\Delta_{0}$ between the two bands (with the indices 1 and 2 correspond to the 3rd and 4th bands respectively). The right hand side of the figure represent the problem in pseudospin space, with $(1h$, $2h)$ as patch-centre and $(1\Lambda$, $2\Lambda)$ as boundaries.} 
\label{Fig:twopatch}
\end{figure}

Above, we have identified the existence of a minimum gap between the third and fourth bands at two points in $k$-space. 
Although the dispersion spectrum in Fig.(\ref{fig:dispersion}) is not symmetric about these two minimum-gap points, we can nevertheless consider the immediate vicinity of these points in constructing an effective two-patch problem. Figure (\ref{Fig:twopatch}) shows the schematic diagram of two-patch problem, where $a$ and $b$ are the two minimum energy-difference patch-centre points in the lower band with momentum indices $k_{1a}$ and $k_{1b}$ respectively. The momentum indices $k_{2a}$ and $k_{2b}$ represent the other two minimum energy-difference patch-centre points in the upper band. These four momenta are connected: $k_{1a}=k_{2a}$ and $k_{1b}=k_{2b}$. In what follows, $\Delta_0$ represents the one-spinon energy gap between two bands, hereafter referred to as the hybridization gap. 
\par
The states present at the patch-centres (labelled by the symbols $(1,2)$ and $(a,b)$) are connected via interband particle-particle(PP) or particle-hole(PH) scattering events between the 3rd and 4th bands. This is induced by the nearest-neighbour (Ising) interaction term in the Hamiltonian \eqref{eqn:Hamiltonian}. In this way, the scattering processes in the vicinity of the two patch-centres form a two patch model described in Fig.{\ref{Fig:twopatch}}. A general pair of electronic states  taken from these two bands are marked by momentum eigenvalues ($\vec{k}_{1a,\Lambda},\vec{k}_{1b,\Lambda-
\delta\Lambda}$) and  ($\vec{k}_{2a,\Lambda'},\vec{k}_{2b,\Lambda'-\delta\Lambda}$), where $\Lambda$ indicates distance from the two patch-centres, and $\delta\Lambda$ the momentum asymmetry within the pair. The two-spinon interaction terms in the Hamiltonian, eq\eqref{eqn:fermionic}, involve the scattering of such pairs of states between the 3rd and 4th bands. These scattering processes come with an additional energy cost above the bare band gap $\Delta_{0}$, $\epsilon_{\delta\Lambda}\propto\delta\Lambda$~. This energy mismatch then leads to a logarithmic singularity in the second order term of the associated T-matrix for the \textit{resonant pairs} chosen symmetrically ($\delta\Lambda =0$) from the two patch-centres
\begin{equation}
T_{1\to 2}(\Delta_0)=(J^{2}\lambda^{2}/\Delta_0)\log(\Delta_0/(\epsilon_{\delta\Lambda}))~.
\end{equation}
Recall that similar log-divergences appear in the calculation of the T matrix for Kondo problem~\cite{phillips2012advanced}, and required a careful RG analysis for further insight. For these symmetrically chosen resonant pair states, there exists PP and PH scattering channels acting on the low energy subspaces: $\hat{n}_{\vec{k}_{1a,\Lambda}}=\hat{n}_{\vec{k}_{1b,\Lambda}}$ and $\hat{n}_{\vec{k}_{1a,\Lambda}}+\hat{n}_{\vec{k}_{1b,\Lambda}}=1$ respectively. The selection procedure for the low energy subspaces employed above is analogous to BCS's construction~\cite{bardeen1957theory,PhysRev.112.1900} used to reach the BCS reduced Hamiltonian starting from a general electronic problem. 
\par
We begin by writing below the effective Hamiltonian formed out of the four patch-centre states $(1/2)a$ and $(1/2)b$ in the PP and PH scattering channels (for a detailed derivation of the Hamiltonian, see Appendix \ref{appendix-I}). In the next section, we will consider the effects of scattering between these patch-centres and the states in their vicinity.
\begin{widetext}
The contribution to the effective Hamiltonian from the PP channel is given by
\begin{eqnarray}
H_{PP} &=&\epsilon_1(\hat{n}_{k1a}+\hat{n}_{k1b})+\epsilon_2(\hat{n}_{k2a}+\hat{n}_{k2b})
+ V_{q=0}\Big(C^{\dagger}_{k1b}C^{\dagger}_{k1a}C_{k1a}C_{k1b}+(1\leftrightarrow 2)\Big)\nonumber\\
&&+ V_{q\neq 0}\Big(C^{\dagger}_{k1b}C^{\dagger}_{k1a}C_{k1a}C_{k1b}+(1\leftrightarrow 2)\Big)
+ V_{q=0}\Big( C^{\dagger}_{k2b}C^{\dagger}_{k2a}C_{k1a}C_{k1b}+(1\leftrightarrow 2)\Big)\nonumber\\
&&+ V_{q\neq 0}\Big(C^{\dagger}_{k2a}C^{\dagger}_{k2b}C_{k1a}C_{k1b}
+(1\leftrightarrow 2)\Big)~.
\label{eqn:HD}
\end{eqnarray}
Similarly, the contribution from the PH channel is
\begin{eqnarray}
H_{PH} &=& V_{q=0}\Big(C^{\dagger}_{k1a}C^{\dagger}_{k2b}C_{k2b}C_{k1a}+(a\leftrightarrow b)\Big)
+ V_{q=0}\Big(C^{\dagger}_{k1a}C^{\dagger}_{k2a}C_{k2a}C_{k1a}+(a\leftrightarrow b)\Big)\nonumber\\
&&+ V_{q\neq 0}\Big(C^{\dagger}_{k2b}C^{\dagger}_{k1a}C_{k2b}C_{k1a}+(a\leftrightarrow b)\Big)
+ V_{q=0}\Big(C^{\dagger}_{k2a}C^{\dagger}_{k1a}C_{k2a}C_{k1a}+(a\leftrightarrow b)\Big)\nonumber\\
&&+ V_{q=0}\Big(C^{\dagger}_{k2b}C^{\dagger}_{k1a}C_{k2a}C_{k1a}+(a\leftrightarrow b)\Big)
+ V_{q\neq 0}\Big(C^{\dagger}_{k1b}C^{\dagger}_{k2b}C_{k2a}C_{k1a}+(a\leftrightarrow b)\Big)\nonumber\\
&&+ V_{q\neq 0}\Big(C^{\dagger}_{k2b}C^{\dagger}_{k1b}C_{k2a}C_{k1a}+(a\leftrightarrow b)\Big)
+ V_{q\neq 0}\Big(C^{\dagger}_{k1b}C^{\dagger}_{k2a}C_{k2b}C_{k1a}+(a\leftrightarrow b)\Big)~.
\label{eqn:spinon}
\end{eqnarray}
\end{widetext}
In $H_{PP}$, $\epsilon_1$ and $\epsilon_2$ are the kinetic energies for the patch-centres in the lower and upper bands respectively. As we take the middle of the two bands as the zero energy surface, $\epsilon_1=-\epsilon_2$. Further,  $\hat{n}_{k1a/b}=C^{\dagger}_{k1a/b}C_{k1a/b}$ and $\hat{n}_{k2a/b}=C^{\dagger}_{k2a/b}C_{k2a/b}$ are the spinon number operators (in $k$-space) at the lower and upper patch-centres respectively. $V_{q=0}$ and $V_{q\neq 0}$ are the zero and non-zero momentum scattering amplitude. In the PH channel, the two spinons scatter from separate bands. Their kinetic energy terms cancel one other, ensuring the absence of kinetic terms from $H_{PH}$. The origin of various scattering terms in the PP and PH channel are described in detail in Appendix \ref{appendix-I}. 
\par 
We define pseudo-spins in the PP channel as follows~\cite{PhysRev.112.1900}, 
\begin{eqnarray}
A^z_{(1/2)D} = \frac{1}{2}(\hat{n}_{k(1/2)b}+\hat{n}_{k(1/2)a}-1),~
A^+_{(1/2)D} = C^{\dagger}_{k(1/2)b} C^{\dagger}_{k(1/2)a}~\text{and}~ A^-_{(1/2)D} = (A^+_{(1/2)D})^\dagger,\label{PP pseudospin}
\end{eqnarray}
and pseudo-spin in the PH channel as
\begin{eqnarray}
A^z_{(1/2)S} =\frac{1}{2}(\hat{n}_{k(1/2)b}-\hat{n}_{k(1/2)a}),~
A^+_{(1/2)S} = C^{\dagger}_{k(1/2)b} C_{k(1/2)a}~\text{and}~
A^-_{(1/2)S} = (A^+_{(1/2)S})^\dagger~,\label{PH pseudospin}
\end{eqnarray}
where the indices $(1/2)$ denote (as above) the 3rd and 4th bands respectively. By first writing the Hamiltonians (\ref{eqn:HD}) and (\ref{eqn:spinon}) in terms of pseudo-spins and then combining both, we have
\begin{eqnarray}
\mathcal{H} = H_{PP}+H_{PH} 
=U(A^z_{2D}+A^z_{1D})+ \Delta (A^z_{2D}-A^z_{1D}) 
&&+2V(\vec{A}_{2D}\vec{A}_{1D}-\vec{A}_{2S}\vec{A}_{1S})~,
\label{twopatchpseudospin}
\end{eqnarray}
where $U=\epsilon_1+\epsilon_2+2V$, $\Delta=\epsilon_2-\epsilon_1 $~. Note that the coupling $V=V_{q=0}-V_{q\neq 0}\equiv \lambda$ (the exchange anisotropy). Further, $V>0$ as $V_{q\neq 0} \propto \cos\Delta k < 0$ (as the momentum transfer is $\pi/2\leq \Delta k\leq 2\pi/3$). 
\section{Renormalization group study for two-patch problem}
\noindent
In the section above, we saw the logarithmic instabilitites in the low energy subspace due to PP or PH scattering processes. Akin to the poor man's scaling approach to the Kondo problem\cite{anderson1970poor}, a treatment of such logarithmic singularities arising at leading order in the coupling $J\lambda$ demands a renormalization group (RG) analysis for the resonant pairs of the two-patch problem. This will enable us in reaching some firm conclusions on the existence and nature of various ground states. Below, we carry out a non-perturbative Hamiltonian RG procedure to treat \textit{quantum fluctuations} of the z-component of the PP and PH pseudospins (eqn\eqref{PP pseudospin}, eqn\eqref{PH pseudospin}) which are induced by pseudospin-flip scattering processes~\cite{mukherjee2018scaling}. The RG procedure involves the iteration of two steps. The first step is the partitioning of the Hamiltonian into a low energy subspace (projection operator $Q_{h}$) involving pseudospins centered at the two patches (Fig.(\ref{Fig:twopatch})), and the boundary subspace (projection operator $Q_{\Lambda}$) with a single pseudospin each located at distances $-\Lambda$ and $\Lambda$ in energy-momentum space. Second, the off-diagonal terms $H^{X}_{h\Lambda}=Q_{\Lambda}HQ_{h}+h.c.$ associated with pseudospin-flip scattering between the boundary ($\Lambda$) and the patch-centre ($h$) states (e.g., terms like $A^{+}_{2D,2\Lambda}A^{-}_{1D,1h}$ + h.c.) are removed via Gauss-Jordan block diagonalization, leading to decoupling of the boundary pseudospin.  Here, $A_{2D,2\Lambda}$ represents the pseudo-spin for upper band in the PP channel with momentum at $\vec{k}_{2\Lambda}$. This RG scheme is similar to Aoki's nonperturbative RG\cite{aoki1982decimation} procedure of Gaussian elimination of single particle states employed for the Anderson disorder problem. The RG is also connected to Glazek and Wilson's nonpertubative Hamiltonian RG procedure~\cite{glazek2004universality} employed for quantum mechanical Hamiltonians, and has been applied recently by some of us to the study of Mott-Hubbard transitions in the 2D Hubbard model~\cite{mukherjee2018scaling}.
\par
The iterative removal of one PP or PH pseudospin at every RG step leads to a RG flow equation for the low-energy two-patch subspace Hamiltonian~
\begin{equation}
H_{h,n-1} = H_{h,n}+H^{X}_{h\Lambda,n}(\omega -H^{D}_{h\Lambda,n})^{-1}H^{X}_{h\Lambda,n}~.
\end{equation}
Here, $H^{X}_{h\Lambda,n}$ is the off-diagonal term coupling states at the energy-momentum boundary (i.e., at $-\Lambda$ and $\Lambda$) and the two patch-centres. The diagonal operator $H^{D}_{h\Lambda,n}$ of the Hamiltonian contains the boundary pseudospin density-density interaction as well a pseudospin dispersion cost. The \textit{quantum fluctuation} scale $\omega$ is the undetermined energy eigenvalue of the system containing contributions from the quantum dynamics of the inter-pseudospin correlations, as well as the pseudospin self-energy induced by the off-diagonal terms. This quantum dynamics is manifested in the Heisenberg equation of motion for the diagonal operator through the non-commutativity of the diagonal and off-diagonal terms.  
\par
By constructing the two-patch problem, we can now write the Hamiltonian in three parts: patch-centre (denoted by $h$), boundaries (denoted by $\Lambda$), and a patch-centre--boundary coupling (denoted as $h\Lambda$) as,
\begin{widetext}
\begin{eqnarray}
H_{h}&=&U(A^z_{2D,2h}+A^z_{1D,1h})+ \Delta (A^z_{2D,2h}-A^z_{1D,1h})
+2V\vec{A}_{2D,2h}\vec{A}_{1D,1h}\nonumber\\
H_{\Lambda}&=&U(A^z_{2D,2\Lambda}+A^z_{1D,1\Lambda})+ \Delta (A^z_{2D,2\Lambda}-A^z_{1D,1\Lambda})
+2V\vec{A}_{2D,2\Lambda}\vec{A}_{1D,1\Lambda}\nonumber\\
H_{h\Lambda} &=& 2V[A^z_{2D,2\Lambda}A^z_{1D,1h}+A^z_{1D,2h}A^z_{1D,1\Lambda}+\frac{1}{2} (A^+_{2D,2\Lambda}A^-_{1D,1h}+\text{h.c.} +A^+_{2D,2h}A^-_{1D,1\Lambda}+\text{h.c.})]~.
\label{eqn:RG1}
\end{eqnarray}
\end{widetext} 
Note that $H_{\Lambda}$ contains an interaction between the two boundaries, while $H_{h}$ contains interactions between the patch-centre states. Further, as the hybridization energy gap term receives no contribution from the PH channel, we will ignore contributions from the PH channel when analysing the effect of interactions in leading to a quantum critical point (i.e., in closing the single particle gap) in section (\ref{hybgaprenorm}). Instead, in section (\ref{2-particle gap}), we will find their contributions as being critical to the opening of a many-body gap beyond the quantum critical point.  
\subsection{Renormalization of hybridization gap}
\label{hybgaprenorm}
The inter-band scattering processes between the patch-centres and the boundaries leads to renormalization of the single particle terms in the {\it two patch} Hamiltonian
\begin{eqnarray}
\Delta H^1_{h} =\frac{V_n^2 A^+_{1D,1h} A^-_{2D,2\Lambda}|\uparrow_{1\Lambda}\uparrow_{2\Lambda}\rangle\langle \uparrow_{1\Lambda}\uparrow_{2\Lambda}|A^+_{2D,2\Lambda} A^-_{1D,1h}}{\langle \uparrow_{1\Lambda}\uparrow_{2\Lambda}|(\omega - H_{\Lambda})|\uparrow_{1\Lambda}\uparrow_{2\Lambda}\rangle}~,
\label{eqn:RG}
\end{eqnarray}
where $\omega$ quantifies the energy cost for pseudospin fluctuations arising from inter-band spin-flip scattering processes.
The renormalisation equation \eqref{eqn:RG} implies that a pseudospin scattering between the lower band patch-centre and the boundary of the upper band involves the excitation to an intermediate configuration  $\uparrow_{1\Lambda}\uparrow_{2\Lambda}$. In the de-excitation process, the pseudospins return to their original configuration from this intermediate configuration (see Fig.(\ref{Fig:twopatch})). We note that the numerator of the flow equations support a one-loop form similar to that obtained from the {\it poor man scaling RG} for the Kondo problem~\cite{anderson1970poor}. However, the denominator contains pseudospin (i.e., two-particle) self energies that are also renormalized in the process. This leads to the non-perturbative nature of the flow equations. We note that a similar feedback in the renormalisation of the two particle vertices is observed in the functional RG formalism as arising from the single-particle self energy~\cite{PhysRevB.86.235140}.    
\par
We now compute the flow equation \eqref{eqn:RG} as follows. We begin by determining the form of the intermediate boundary pseudospin state propagator, and then compute the operation of the spin flip terms on the pseudospin state space. The zero energy surface lies exactly in the middle of the two bands, such that the resonant PP pair possesses zero net kinetic energy,  $\epsilon_{1\Lambda}=-\epsilon_{2\Lambda}$. The coefficient of the effective field-like term in eqn.\eqref{eqn:RG1} simplifies to $U_0=2V_0$, where $V_0$ is the bare interaction strength. $V_n$ is the renormalised interaction strength/two-particle self-energy obtained from the $n$th step of the RG process. The intermediate boundary state's eigenenergy is given by, 
\begin{equation}
H_{\Lambda}|\uparrow_{1\Lambda}\uparrow_{2\Lambda}\rangle=(U_0+\frac{V_n}{2})|\uparrow_{1\Lambda}\uparrow_{2\Lambda}\rangle~.
\end{equation}
The action of the pseudospin-flip operators on the intermediate state 
\begin{eqnarray}
A^-_{2D,2\Lambda}|\uparrow_{1\Lambda}\uparrow_{2\Lambda}\rangle=|\uparrow_{1\Lambda}\downarrow_{2\Lambda}\rangle
\end{eqnarray}
allows determination of the numerator in eqn\eqref{eqn:RG}. Thus, the renormalisation equation takes the form
\begin{eqnarray}
\Delta H^1_{h} &=&\frac{V_n^2 A^+_{1D,1h}A^-_{1D,1h}|\uparrow_{1\Lambda}\downarrow_{2\Lambda}\rangle\langle \uparrow_{1\Lambda}\downarrow_{2\Lambda}|}{\omega-(2V_0+\frac{V_n}{2})}\nonumber \\
&=&\frac{V_n^2 (\frac{1}{2}+A^z_{1D,1h})|\uparrow_{1\Lambda}\downarrow_{2\Lambda}\rangle\langle \uparrow_{1\Lambda}\downarrow_{2\Lambda}|}{\omega-(2V_0+\frac{V_n}{2})}~.
\end{eqnarray}
Similarly, the pseudospin scattering between upper band patch-centre and the boundary of the lower band leads to
\begin{eqnarray}
\Delta H^2_{h} &=&\frac{V_n^2 A^+_{2D,2h} A^-_{1D,1\Lambda}|\uparrow_{1\Lambda}\uparrow_{2\Lambda}\rangle\langle \uparrow_{1\Lambda}\uparrow_{2\Lambda}|A^+_{1D,1\Lambda} A^-_{2D,2h}}{\langle \uparrow_{1\Lambda}\uparrow_{2\Lambda}|(\omega - H_{\Lambda})|\uparrow_{1\Lambda}\uparrow_{2\Lambda}\rangle}\nonumber\\
&&=\frac{V_n^2 (\frac{1}{2}+A^z_{2D,2h})|\downarrow_{1\Lambda}\uparrow_{2\Lambda}\rangle\langle \downarrow_{1\Lambda}\uparrow_{2\Lambda}|}{\omega-(2V_0+\frac{V_n}{2})}~.
\end{eqnarray}
Together, the complete contribution to the Hamiltonian renormalization from the above processes is given by 
\begin{widetext}
\begin{eqnarray}
\Delta H^1_{h}+\Delta H^2_{h} &=& \frac{V_n^2}{\omega-(2V_0+\frac{V_n}{2})}(A^z_{1D,1h}|\uparrow_{1\Lambda}\downarrow_{2\Lambda}\rangle\langle \uparrow_{1\Lambda}\downarrow_{2\Lambda}|
+ A^z_{2D,2h}|\downarrow_{1\Lambda}\uparrow_{2\Lambda}\rangle\langle \downarrow_{1\Lambda}\uparrow_{2\Lambda}|)\nonumber \\
&=& \frac{K}{2}[(A^z_{2D,2h}-A^z_{1D,1h})(|\downarrow_{1\Lambda}\uparrow_{2\Lambda}\rangle\langle \downarrow_{1\Lambda}\uparrow_{2\Lambda}|-|\uparrow_{1\Lambda}\downarrow_{2\Lambda} \rangle\langle \uparrow_{1\Lambda}\downarrow_{2\Lambda}|)\nonumber \\
&&+(A^z_{2D,2h}+A^z_{1D,1h})(|\downarrow_{1\Lambda}\uparrow_{2\Lambda}\rangle\langle \downarrow_{1\Lambda}\uparrow_{2\Lambda}|+|\uparrow_{1\Lambda}\downarrow_{2\Lambda} \rangle\langle \uparrow_{1\Lambda}\downarrow_{2\Lambda}|)]
\label{eqn:RGhybridization}~,
\end{eqnarray}
\end{widetext}
where $K=\frac{V_n^2}{\omega-(2V_0+\frac{V_n}{2})}$. By taking a trace of Eqn.(\ref{eqn:RGhybridization}) with respect to the final pseudospin state $|\downarrow_{1\Lambda}\uparrow_{2\Lambda}\rangle$, the renormalized hybridisation term of the patch-centre Hamiltonian ($H_{h}$ in Eqn\eqref{eqn:RG1}) attains the form 
\begin{eqnarray}
(\Delta+\frac{K}{2})(A^z_{2D,2h}-A^z_{1D,1h})~.
\end{eqnarray}
This leads to the RG flow equation for the hybridization gap $\delta\Delta_n=\Delta_{n+1}-\Delta_n$ and the interaction strength $\delta V_n = V_{n+1} - V_n$ as (for details see Appendix-\ref{appendix-II})
\begin{eqnarray}
\delta \Delta_n=\frac{K}{2}=\frac{2~(V_n/2)^2}{c|\omega-2V_0|-\frac{V_n}{2}}\equiv \delta V_n~,
\label{eqn:hybRg}
\end{eqnarray} 
where $c =\text{sgn}(\omega -2V_0)$ is the sign of the non interacting two particle Green function.
\par
\begin{figure}[h!]
\centering
\includegraphics[scale=.35]{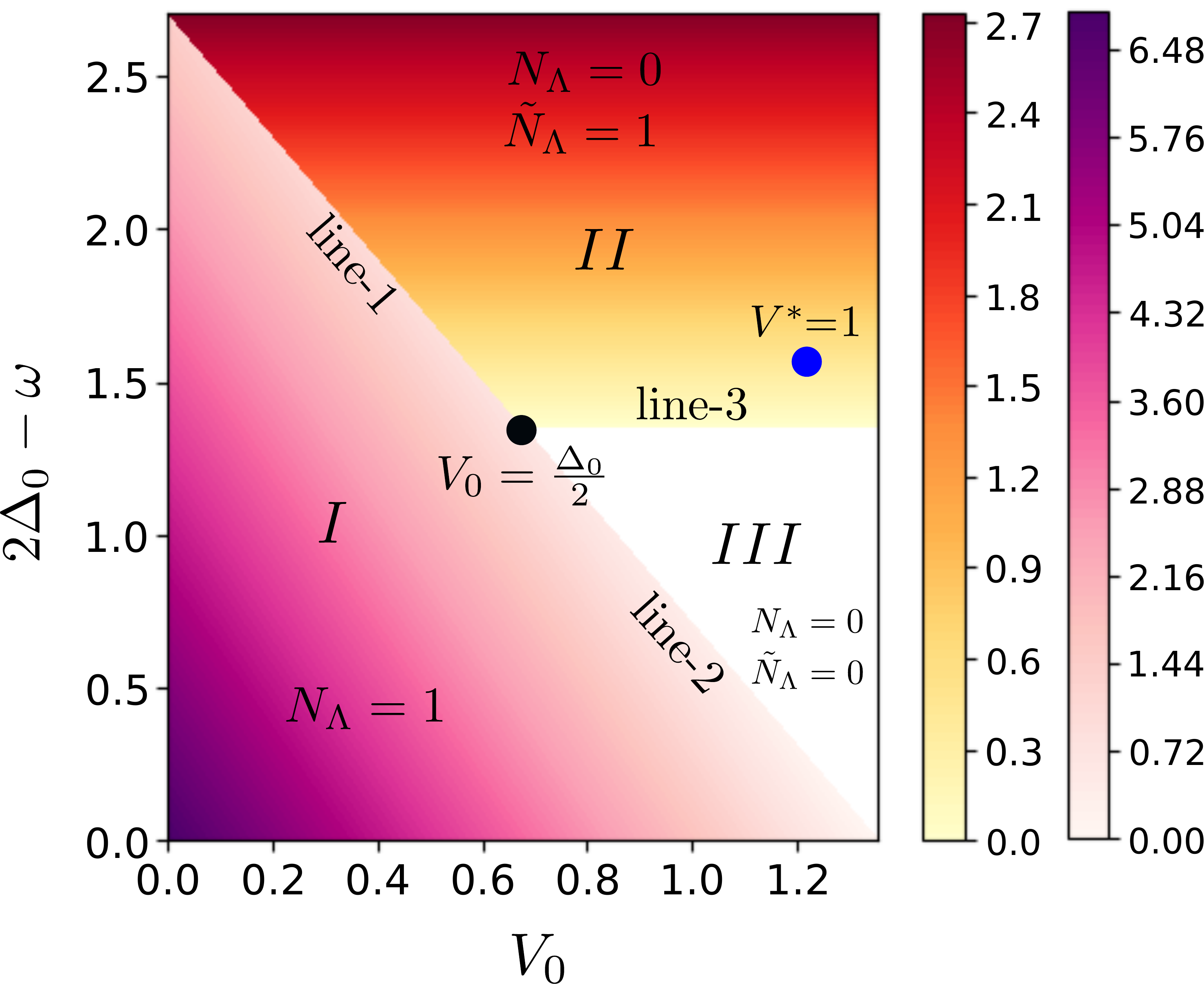}
\caption{RG phase diagram for n.n. $S=1/2$ kagome antiferromagnet, representing the one-spinon gap (pink bar) and two-spinon gap (orange-yellow bar) magnitudes as a function of the bare interaction $V_0 (\equiv\lambda$,  the exchange anisotropy) and the effective energyscale for quantum fluctuations, $2\Delta_0-\omega$. The region $I$ (pink) represents a one-spinon gapped topological spin liquid phase, while region $II$ (orange-yellow) represents a two-spinon gapped topological spin liquid phase. Region $III$ (white) represents the gapless algebraic spin liquid. The black circle ($V_0=\Delta_0/2, \omega=\Delta_0$) marks the multicritical point lying at the meeting of three critical lines (1, 2 and 3). The blue circle ($V^{\ast}=1$) corresponds to the Heisenberg point.} 
\label{fig:phasediagram}
\end{figure}

In the RG equation given above, for $\omega=\Delta_{0}$, the quantity $c$ will change sign from $+Ve$ to $-Ve$ for $\Delta_{0}<2V_{0}$. This makes the RG flows for both the hybridization gap and inter-band scattering term irrelevant. This marks  $\Delta_{0}/2=V_0^{*}=0.68$ (in units of $J$) as a $SU(2)$-symmetric quantum critical point across which the hybridization gap-closing transition takes place, revealing a singular Fermi surface with two Dirac points. This is shown in Fig.(\ref{fig:phasediagram}) as a black filled circle. The critical value $V_{0}^{*}=0.68$ obtained is close to the mean-field value of $0.6$ obtained by Kumar et al. in Ref.(\cite{PhysRevB.90.174409}). As we will see in subsection \ref{gaplesssubsection} below, this special point lies at the meeting of three quantum critical lines, two of which possess $SU(2)$-symmetry (line-2 and line-3 in Fig.(\ref{fig:phasediagram})) and the third a $U(1)$ symmetry (line-1 in Fig.(\ref{fig:phasediagram})). Indeed, this point corresponds to a multicritical point as it is reached at a special value of $\Delta_{0}$, $V_{0}$ and $\omega$.
\par
On the other hand, in the regime $V_0<\frac{\Delta_0}{2}$ (marked as $I$ in Fig.(\ref{fig:phasediagram})), the interband scattering terms assists the hybridization gap and both the RG flows are relevant (as can be seen from eqn.(\ref{eqn:hybRg})). Here, stable fixed point values are reached for the scattering coupling $V_n^\ast=2(2V_0-\omega)$ and the hybridization gap $\Delta_{n}^{\ast}= \Delta_{0}+2\omega -5V_{0}$~(see Appendix \ref{appendix-II}). The effective two patch-centre Hamiltonian at this stable fixed point gives the low-energy theory of the $1/3$rd plateau
\begin{eqnarray}
H^\ast(\omega) = 2V_{0}(A^{z}_{2D,2h}+A^{z}_{1D,1h})+\Delta^{*}(\omega)(A^z_{2D,2h}-A^z_{1D,1h})
+V^{*}(\omega)\vec{A}_{2D,2h}\vec{A}_{1D,1h}~.
\end{eqnarray}
\par
We can now detail some of the properties of the gapped ground state on this magnetisation plateau. Following Ref.(\cite{pal2018non}), applying the twist operator 
\begin{eqnarray}
\hat{O} = \hspace*{-0.05cm}\exp\hspace*{-0.05cm} \big[i\frac{2\pi}{N_1}\Big(\hspace*{-0.1cm}\sum_{\vec{r}}\hspace*{-0.1cm}(n_1\hspace*{-0.1cm}+\hspace*{-0.1cm}\frac{n_2}{2}) \hat{S}^z_{\vec{r}}+\hspace*{-0.1cm}\sum_{\vec{R}}(\frac{1}{4} \hat{S}^z_{\vec{R},b}\hspace*{-0.1cm}+\hspace*{-0.1cm}\frac{1}{2} \hat{S}^z_{\vec{R},c})\Big)\big]
\label{eqn:twist}
\end{eqnarray}
twice on the ground state wave function returns the ground state wave function to itself, yielding a gapped twofold ground state degeneracy on the torus~\cite{PhysRevB.65.153110,PhysRevLett.84.1535,PhysRevLett.90.236401}. This gap opening is an outcome of the two particle scattering across the two-patch centres (see Fig.(\ref{Fig:twopatch})). This scattering process is described by the pseudospin-flip part of the Hamiltonian eqn(\ref{twopatchpseudospin}), and can be connected to the PP-projected twist operator as follows (for details see Appendix \ref{appendix-IV})
\begin{eqnarray}
\mathcal{H}_D^{\pm}= V^{*}(\omega)[A^+_{2D}A^-_{1D}+h.c.]= V^{*}(\omega) P_{D}(\hat{O}^2+\hat{O}^{\dagger 2})P_{D}~.
\label{eqn:twistscattering}
\end{eqnarray}
where $\mathcal{H}_D^{\pm}$ is the PP channel's spin-flip part of the Hamiltonian (\ref{twopatchpseudospin}), and $P_{D}=(16/9)\vec{A}^{2}_{1D}\vec{A}^{2}_{2D}$ is the projection operator on the PP channel for the 3rd and the 4th bands. In order to probe the topology of the ground state manifold, we define the center of mass translation operator $\hat{T} = \exp[iP_{cm}]$ (where $P_{cm}$ is the total momentum). From the non-commutativity of the twist and translation operators we obtain a $Z_{2}$ invariant 
\begin{eqnarray}
P_{D}\hat{T}_{\hat{a}_1}\hat{O}\hat{T}^\dagger_{\hat{a}_1}\hat{O}^{\dagger}P_{D}&=& \exp\big[-i\pi\big]~,\nonumber\\
P_{D}\hat{T}_{\hat{a}_1}\hat{O}^{2}\hat{T}^\dagger_{\hat{a}_1}\hat{O}^{2\dagger}P_{D}&=& \exp\big[-i2\pi\big]~.
\label{LSMsinglet}
\end{eqnarray}
This $Z_{2}$ invariant indicates that the states $|\psi_0\rangle$ and $\hat{O}|\psi_0\rangle \equiv|\psi_{1}\rangle$ are orthogonal to one another ($\langle \psi_{1}|\psi_{0}\rangle =0 $) and comprise the ground state manifold. Further, that $\hat{O}^{2}|\psi_0\rangle=|\psi_{0}\rangle$~. The topological phase $\pi$ that arises out of this non-commutativity allows us to extract a fractional charge $q = \frac{1}{2}$ associated with the spectral flow between the two ground states~\cite{PhysRevLett.96.060601,PhysRevB.97.115138}. Further, we will also see in section (\ref{hallsection}) below that the many-body gap originating from the operator $\hat{O}^{2}$ protects the degenerate ground state manifold, resulting in a topological Chern number associated with a spin Hall conductivity. Our findings are consistent with the Chern-Simons gauge field theory developed in Kumar et al~\cite{PhysRevB.90.174409}.
%====================================================================================================
\subsection{Gap opening beyond the Quantum critical point}\label{2-particle gap}
We will study the putative gapped phase lying beyond the quantum critical point ($V_0>\frac{\Delta_0}{2}$) induced by scattering in the PP/PH channels (marked as region $II$ in Fig.(\ref{fig:phasediagram})). To do so, we begin with the effective Hamiltonian at the band touching points involving both the PP and PH pseudospins comprised of Dirac electrons present in the neighbourhood of the two Dirac points
\begin{widetext}
\begin{eqnarray}
H_{h} = 2(V_0-\frac{\Delta_0}{2})(A^z_{2D,2h}+A^z_{1D,1h})+2V_0[\vec{A}_{2D,2h}\cdot\vec{A}_{1D,1h}-
\vec{A}_{1S,1h}\cdot\vec{A}_{2S,2h}]~,\nonumber\\
H_{\Lambda} = 2(V_0-\frac{\Delta_0}{2})(A^z_{2D,2\Lambda}+A^z_{1D,1\Lambda})+2V_0[\vec{A}_{2D,2\Lambda}\cdot\vec{A}_{1D,1\Lambda}
-\vec{A}_{2S,2\Lambda}\cdot\vec{A}_{1S,1\Lambda}]~, 
\nonumber\\
H_{h\Lambda} = 2V_0[(\vec{A}_{2D,2\Lambda}\cdot\vec{A}_{1D,1h}+\vec{A}_{1D,1\Lambda}\cdot\vec{A}_{2D,2h})
-(\vec{A}_{2S,2\Lambda}\cdot\vec{A}_{1S,1h}+ \vec{A}_{1S,1\Lambda}\cdot\vec{A}_{1S,2h})]~.
\label{spinon_ham}\end{eqnarray}
\end{widetext}
As the single particle gap $\Delta_{0}$ in Fig. (\ref{Fig:twopatch}) vanishes, the lowest energy electrons have their energy shifted by $\frac{1}{2}\Delta_{0}$; zero energy electrons are now present at the Dirac points (i.e., in the middle of two renormalized bands). This energy shift is accounted for by offsetting the flat dispersion of the PP pseudospins (first term of $H_{\Lambda}$ in eqn\eqref{spinon_ham}) by the same amount. Now considering intermediate state for PP and PH channel, $|\psi_{\Lambda}\rangle_{D} = |\uparrow_{1\Lambda}\uparrow_{2\Lambda}\rangle_D$ and 
$|\psi_{\Lambda}\rangle_{S} = |\uparrow_{1\Lambda}\uparrow_{2\Lambda}\rangle_S$ respectively, the corresponding boundary energies are computed from 
%\textcolor{blue}{ $V_n$ in the below will be $V_n^D$ and $V_n^S$}.
\begin{eqnarray}
\langle\psi_{\Lambda}|H_\Lambda|\psi_{\Lambda}\rangle_{D} = 2(V_0-\frac{\Delta_0}{2})+\frac{V_{n}^{D}}{2},~~ \langle\psi_{\Lambda}|H_\Lambda|\psi_{\Lambda}\rangle_{S} = -\frac{V_{n}^{S}}{2}~. 
\end{eqnarray}
The RG equation for PP/PH channels are then found to be
\begin{eqnarray}
V^D_{n+1}-V^D_n &=& \frac{(V^D_n)^2}{(\omega-\Delta_0)-2(V_0-\frac{\Delta_0}{2})-(\frac{V_n^D}{2})}~, \nonumber \\
V^S_{n+1}-V^S_n &=& -\frac{(V^S_n)^2}{(\omega-\Delta_0)+(\frac{V_n^S}{2})}~,
\label{eqn:RGcritical}
\end{eqnarray}
where the bare values of the couplings $V_0^D=V_0^S=V_0$. It is worth noting that, due to the closure of the hybridisation gap, the effective energy scale for quantum fluctuations is now given by $(\omega - \Delta_{0})$. From the RG equations eqn\eqref{eqn:RGcritical}, we observe that in the regime $V_{0}>\Delta_{0}-\frac{1}{2}\omega$, the PP coupling $V^{D}$ is RG irrelevant whereas the PH coupling $V^{S}$ is RG relevant. The relevant RG flow stops at the finite fixed point value $V^{S\ast}=2(\Delta_0-\omega)$ as a function of $\omega$ and $\Delta_{0}$. Thus the latter forms a gap in the PH pseudospin Hilbert space, and determines the resultant stable fixed point Hamiltonian 
\begin{equation}
H^{*}_{h} = -2V^{S*}\vec{A}_{1S,1h}\cdot\vec{A}_{2S,2h}~.
\label{2particleeffham}
\end{equation}
Similar to Eqn.(\ref{eqn:twistscattering}) for PH interband scattering channel, one can show that the spin-flip part in the fixed point Hamiltonian, 
\begin{eqnarray}
\mathcal{H}_S^{\pm}= -V^{S\ast}[A^+_{2S}A^-_{1S}+h.c.]= -V^{S\ast}P_S(\hat{O}^2+\hat{O}^{\dagger 2})P_S
\label{eqn:PHtwist}
\end{eqnarray}
can be written in terms of the doubled twist operator projected onto the PH pseudospin space (see Appendix \ref{appendix-IV} for details). This allows us, once again, to predict a gapped, two-fold degenerate ground state on the torus~\cite{PhysRevB.65.153110}. Further, following the arguments presented earlier, spectral flow between the two degenerate ground states is associated with a fractional charge of $q=\frac{1}{2}$.
\par
In order to place the kagome Heisenberg antiferromagnet in the presence of an external magnetic field along the $z$-direction (\eqref{eqn:Hamiltonian}) with $\lambda=1$, $h\neq 0$) in the RG phase diagram, we recall that $\lambda\equiv V$ in our formalism. Thus, the stable fixed point $V^{S\ast}=1$ corresponds to $\lambda^{*}=1$. This gives $2\Delta_0-\omega=\Delta_0+\frac{1}{2}$. This places the Heisenberg point in region $II$ of the phase diagram at the point $(1, \Delta_0+\frac{1}{2})$, as indicated by blue filled circle in Fig.(\ref{fig:phasediagram}). As indicated by the effective Hamiltonian in eqn.(\ref{2particleeffham}) above, the two-particle gapped phase corresponds to a gapped $SU(2)$ symmetric phase comprised of electron-hole pairs from the 3rd and 4th bands. This effective Hamiltonian was reached by an RG analysis respecting the $SU(2)$ symmetry of quantum fluctuation terms (eqns.\eqref{eqn:RGcritical}) arising from the non-commutativity of various terms in eqn\eqref{spinon_ham} (e.g., $[\vec{A}^{s}_{2\Lambda}\cdot\vec{A}^{s}_{1h},\vec{A}^{s}_{2h}\cdot\vec{A}^{s}_{1h}]\neq 0$). This indicates that we could have equivalently studied a $SU(2)$ non-Abelian lattice gauge theory on the kagome lattice  associated  with such quantum fluctuations~\cite{RevModPhys.51.659}. In the continuum limit, such a $SU(2)$-symmetric non-abelian gauge theory possesses a disordered ground state with a dynamically generated mass gap. Such a gauge theory can be obtained from a fermionic non-linear sigma model of massive Dirac fermions in $(2+1)$ dimensions coupled to a SU(2) order parameter~\cite{abanov2000theta}, and possesses a topological Hopf term. We will further quantify the topological invariants of the gapped phases in regions $I$ and $II$ in the next section. 
\subsection{Gapless phase and phase boundaries}
\label{gaplesssubsection} 
In the regime $V_{0}>\Delta_{0}-\frac{1}{2}\omega$, both PP and PH scattering processes are irrelevant, leading to a state with robust gapless Dirac spinons. This is indicated as region $III$ of the phase diagram in Fig.(\ref{fig:phasediagram}), and corresponds to an algebraic spin liquid \cite{PhysRevB.77.224413,PhysRevX.7.031020}. Here, the RG flows in eqn.(\ref{eqn:RGcritical}) stop at the fixed point values of vanishing interaction strength $V_n^{\ast}=0$ and hybridisation gap $\Delta_{n}^{*}=0$. This corresponds to the appearance of the Dirac point Fermi surface, with a vanishing effective patch-centre Hamiltonian, $H_{III}=0$. We will further elucidate the properties of this gapless phase in section (\ref{gaplesssection}).
\par 
Finally, we can define fixed point Hamiltonians for the various phase boundaries in the phase diagram. The fixed point Hamiltonian for the boundary separating Phase-I from phase-II (line-1 in Fig.(\ref{fig:phasediagram})) is 
\begin{equation}
H_1=(V_{0}-\frac{\Delta_{0}}{2})(A^z_{1D}+A^z_{2D})~,
\end{equation}
indicating that a gapless Dirac-point Fermi surface is revealed along 1, but with a shift in the effective chemical potential to the value $V_{0}-\frac{\Delta_{0}}{2}$. $H_{1}$ thus corresponds to a line of quantum critical theories with $U(1)$-symmetry, and possesses a topological theta-term~\cite{fradkin2013field}. On the other hand, for the boundaries separating phase-I from Phase-III (line-2) and that separating phase-II from phase-III (line-3) corresponding to $SU(2)$-symmetric {\it critical theories}: $H_2=H_3=0$. The vanishing effective Hamiltonian for such gapless Dirac cones on line-2, line-3 and phase $III$ indicates an emergent particle-hole symmetry. This leads to a $SU(2)$-symmetric Wess-Zumino-Novikov-Witten (WZNW) topological term with coefficient $\tilde{S}=1/2$~\cite{fradkin2013field} in the theory. 
%===========================================================================================
\section{Topological quantum numbers and spin Hall conductivity}
\label{hallsection}
We now use various topological quantum numbers to distinguish the different phases in the phase diagram Fig.(\ref{fig:phasediagram}). We begin by rewriting the RG eqn.(\ref{eqn:hybRg}) as
\begin{eqnarray}
\delta \Delta_n &=& \frac{2(V_n/2)^2}{\text{Sgn}(G^{-1}_{0,\Lambda})~|G^{-1}_{0,\Lambda}|-\frac{V_n}{2}}~,
\end{eqnarray}
where $G^{-1}_{0,\Lambda}(\omega)= \omega-2V_0$, such that $\text{Sgn}(G^{-1}_{0,\Lambda}(\Delta_0))=\text{Sgn}(1-\frac{2V_0}{\Delta_0})$. Thus, the quantity $\text{Sgn}(G^{-1}_{0,\Lambda}(\Delta_0))$ decides whether the RG equation for the hybridisation gap $\Delta$ is relevant (+ve) or irrelevant (-ve).
\par
We now define the complex function,
\begin{eqnarray}
\tilde{G}_0(z)=\frac{1}{z-G_{0,\Lambda}^{-1}(\Delta_0)}~,
\label{eqn:complexG}
\end{eqnarray}
and the corresponding topological index~\cite{unruh2007quantum,volovik2003universe} as
\begin{eqnarray}
N_\Lambda &=& \frac{1}{2\pi i}\int dz~ \tilde{G}_0(z)~.
\label{eqn:topologinumber}
\end{eqnarray}
The value of $N_\Lambda$ is then evaluated as (for details see Appendix-\ref{appendix-III})
\begin{eqnarray}
N_\Lambda &=& 1 \qquad \text{if} \quad \Delta_0> 2V_0 \nonumber \\
 &=& 0 \qquad \text{if} \quad \Delta_0< 2V_0~,
\end{eqnarray}
indicating that $N_\Lambda$ is non-trivial in phase $I$ (the one-spinon gapped phase) and trivial otherwise.
\par
Similarly, we can define a topological index from the RG eqn.(\ref{eqn:RGcritical}) as follows
\begin{eqnarray}
\tilde{N}_{\Lambda}=\frac{1}{2\pi i}\int dz~ \tilde{G}^S_0(z)~,
\end{eqnarray}
where $\tilde{G}^S_0(z)=1/(z-G_{0,\Lambda}^{-1}(\omega)_S)$ and $G_{0,\Lambda}^{-1}(\omega)_S=\omega-\Delta_0$. We then find that $\tilde{N}_{\Lambda}$ is non-trivial in phase $II$ (the two-spinon gapped phase) and trivial otherwise.
\begin{eqnarray}
\tilde{N}_{\Lambda} &=& 1 \quad \text{if} \quad  \omega>\Delta_0 \nonumber \\
 &=& 0 \quad \text{if} \quad  \omega<\Delta_0 
\end{eqnarray}
\par
Finally, by mapping the problem in phase $I$ close to the gap-closing point to that of massive Dirac spinons, we find the Chern number $C=1$ (see Appendix-\ref{appendix-III}). Then, following Kumar \emph{et al.}~\cite{PhysRevB.90.174409}, we can connect this Chern number with an  induced Chern-Simons term in the corresponding gauge-field theory ($\theta_F$), $\theta_F = C/2\pi = 1/2\pi$. Together with the original Chern-Simons statistical angle $\theta=1/2\pi$, this leads to an effective Chern-Simons coupling~\cite{PhysRevB.90.174409}
\begin{eqnarray}
\theta_{eff}= \frac{\theta~\theta_{F}}{\theta + \theta_{F}} = \frac{1}{2}\frac{1}{2\pi}~,
\end{eqnarray}
and, thence, the fractional spin hall conductivity
\begin{eqnarray}
\sigma_{xy}^s= \theta_{eff}=\frac{1}{2}\frac{1}{2\pi}=\frac{\nu}{2\pi}~.
\end{eqnarray}
\section{Properties of gapless phase in the RG phase diagram}
\label{gaplesssection}
Finally, we turn to a detailed analysis of Region III in the phase diagram Fig \ref{fig:phasediagram}. As discussed above, we find here a gapless phase with Dirac spinons, constituting a singular Fermi surface of two Dirac-points. The PP and PH scattering processes that participate in gap opening were earlier found to be RG irrelevant in this phase (see Sec. \ref{2-particle gap}). Nevertheless, the velocity of the Dirac spinons can well undergo a nontrivial renormalization due to inter-particle interaction. In turn, such a velocity renormalization will affect the spinon self-energy ($\Sigma$), quasiparticle residue ($Z$) and lifetime ($\tau$) of the gapless spinons. An investigation of these properties is presented below.
\begin{figure}[h!]
\centering
\includegraphics[scale=0.9]{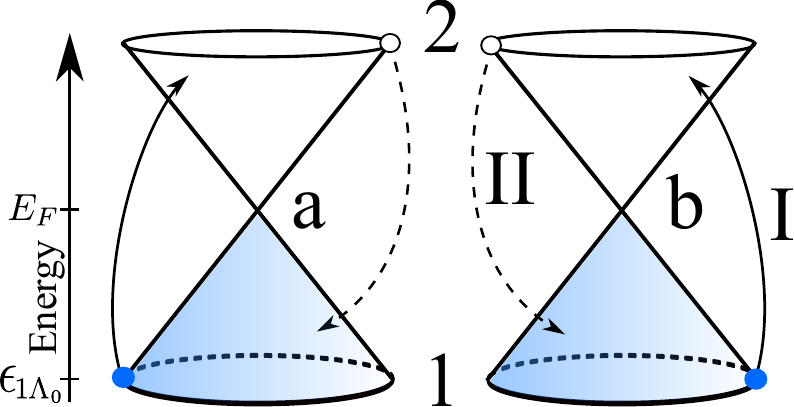}
\caption{Phase-III, with the 3rd and 4th bands (denoted by the symbols 1 and 2) touching at two Dirac points (shown as a and b). The lower (blue) band is filled, with the Fermi energy $E_{F}$ lying at the Dirac points. Here, particles/holes (solid blue/empty circles) from lower/upper band can scatter to upper/lower band via the processes I/II (indicated via solid/dotted arrows). The solid blue circles indicate the highest energy spinons in the lower band, with $\epsilon_{1\Lambda_{0}}$ being the high energy cutoff.}
\label{fig:DiracRG}
\end{figure}
\par
To begin with, we define the energy bandwidth in the problem as $W = \epsilon_{1\Lambda_{0}}-\epsilon_{2\Lambda_{0}}=2\epsilon_{1\Lambda_{0}}$, as $\epsilon_{2\Lambda_{0}}= -\epsilon_{1\Lambda_{0}}$ due to the emergent particle-hole symmetry in this phase. $W$ is the energy difference between the highest energy electrons/holes placed in the lower/upper boundaries of the Dirac cones (see Fig.(\ref{fig:DiracRG})). The renormalization of the spinon dispersion (and hence the spinon velocity) takes place via the scattering of the pair of spinons (from the a and b Dirac cones in Fig.(\ref{fig:DiracRG})) from the lower Dirac cone boundary $\epsilon_{1\Lambda_{0}}$ to the upper Dirac cone window ($\Lambda <\Lambda_{0}$). The excitation process is via the following term in the Hamiltonian, $V_{0}A^{+}_{2D,2\Lambda}A^{-}_{1D,1\Lambda_{0}}$ (shown as process I in Fig.(\ref{fig:DiracRG})), followed by de-excitation to the lower Dirac cone, ($V_{0}A^{+}_{1D,1\Lambda_{0}}A^{-}_{2D,2\Lambda}$). The other process (indicated as II in Fig.(\ref{fig:DiracRG}), via dotted arrows) proceeds in the same way, but for the holes in the upper Dirac cone. The RG flow proceeds as the bandwidth is lowered $\epsilon_{1\Lambda_{0}}>\epsilon_{1\Lambda_{1}}>\ldots$ from one step to the next. The Hamiltonian for the states lying within the cut-off scale $\Lambda_{n}$ for the lower Dirac cone has the form
\begin{eqnarray}
&& H_{1,n}= \epsilon_{1\Lambda_n}(\hat{n}_{\vec{k}_{\Lambda_n,1a}}+\hat{n}_{\vec{k}_{\Lambda_n,1b}})+V\hat{n}_{\vec{k}_{\Lambda_n,1a}} \hat{n}_{\vec{k}_{\Lambda_n,1b}}\nonumber \\
&&= (2\epsilon_{1\Lambda_n}+V)A^{z}_{1D,\Lambda_n}+V\big[(A^{z}_{1D,\Lambda_n})^2- (A^{z}_{1S,\Lambda_n})^2\big] \nonumber \\
&&\simeq (2\epsilon_{1\Lambda_n}+V) A^{z}_{1D,\Lambda_n}~,
\end{eqnarray}
where $V$ is the interaction strength as defined in earlier section, and we have neglected certain constant terms ($\propto (A^{z})^{2}$). The Hamiltonian $H_{1,n}$ has spinon kinetic energy terms and correlation energy terms that preserve the gapless nature of the spectrum in this phase. As the RG proceeds by lowering the bandwidth, $\epsilon_{1\Lambda_n}$ is the spinon dispersion generated at the nth RG step for states placed at the renormalized lower Dirac cone boundary.  
\par
The $A$'s are pseudospin operators (see Appendix \ref{appendix-I}) defined in the associated Hilbert space of fermions placed at the boundary. The structure of the effective Hamiltonian $H_{1,n}$ above is reminiscent of the Fermi liquid, i.e., it contains a kinetic energy linear in $n_{\vec{k}}$ and inter-particle interactions that are quadratic in in $n_{\vec{k}}$. Further, $ \hat{n}_{\vec{k}_{\Lambda_n,1a}} -1/2 = A^{z}_{1D,\Lambda_n}$ is conserved during the RG transformations. The associated Greens function for pseudospin $A_{1D,\Lambda_{n}}$ in an intermediate configuration $\downarrow$ at renormalised lower Dirac cone boundary $\Lambda_{n}$ (i.e., with energy $\epsilon_{1\Lambda_n}$) is
\begin{eqnarray}
G_{1,n,\downarrow}(\omega_{1})=\frac{|\downarrow_{1\Lambda_n}\rangle\langle \downarrow_{1\Lambda_n}|}{\omega_1+\epsilon_{1\Lambda_n}+\frac{V_0}{2}}~,
\end{eqnarray}
where $\omega_1$ is the scale for quantum fluctuations of the spinon dispersion due to two-spinon interactions. Then, the RG equation for process-II in Fig.(\ref{fig:DiracRG}) is
\begin{eqnarray}
\epsilon_{2\Lambda}^{n-1}-\epsilon_{2\Lambda}^n &=& -\frac{V_0^2}{2(\omega_1+\epsilon_{1\Lambda_n}+\frac{V_0}{2})}~,
\label{eqn:DiracRG1}
\end{eqnarray}
where $\epsilon_{2\Lambda}$'s are the dispersion for the spinons in the upper Dirac cone (upper band).
Similarly for process-I,
\begin{eqnarray}
\epsilon_{1\Lambda}^{n-1}-\epsilon_{1\Lambda}^n &=& \frac{V_0^2}{2(\omega_2-\epsilon_{2\Lambda_n} -\frac{V_0}{2})}
\label{eqn:DiracRG2}
\end{eqnarray}
where, $\epsilon_{2\Lambda_n}$ is the fermion dispersion at the renormalized upper Dirac cone boundary, the $\epsilon_{1\Lambda}$'s are the dispersion in the lower band and $\omega_2$ is the fluctuation energyscale for the process (for a detailed derivation, see Appendix \ref{appendix-III}).  %As the Dirac spectrum is symmetric about band touching Dirac points, one would expect that  $|\epsilon_{1\Lambda}|=-|\epsilon_{2\Lambda}|$. 
Putting the  constraint $|\epsilon_{1\Lambda}|=-|\epsilon_{2\Lambda}|$ (particle-hole symmetry of the massless Dirac spectrum) on  eqn.(\ref{eqn:DiracRG2}), we have
\begin{eqnarray}
\epsilon_{2\Lambda}^{n-1}-\epsilon_{2\Lambda}^n = -\frac{V_0^2}{2(\omega_2+|\epsilon_{1\Lambda_n}| -\frac{V_0}{2})}~.
\label{eqn:DiracRG11}
\end{eqnarray}
Comparing eqns. (\ref{eqn:DiracRG1}) and (\ref{eqn:DiracRG11}), we obtain a relation between the two quantum fluctuation energyscales: $\omega_2=V_0+\omega_1$. Using this to simplify eqn.(\ref{eqn:DiracRG2}), we obtain
\begin{eqnarray}
\epsilon_{1\Lambda}^{n-1}-\epsilon_{1\Lambda}^n = \frac{V_0^2}{2(\omega_1+ |\epsilon_{1\Lambda_n}| +\frac{V_0}{2})}~.
\label{eqn:DiracRGF}
\end{eqnarray}
The energyscales $\omega_{1}$ and $\omega_{2}$ are associated with the fluctuations of the z-component of a single pseudospin in the lower and upper band respectively. The RG equations show that they renormalize the pseudospin dispersion relation, eqns. \eqref{eqn:DiracRG11} and \eqref{eqn:DiracRGF}. Recall that in the RG relation for the stability of the gapped phases I and II, the intermediate configuration energy involved the interaction cost of a pseudospin from the lower band and another from upper band (eqn\eqref{eqn:RG1}). Thus, the fluctuation scale $\omega$ for the RG gapped phases is double that of the scale in the gapless phase: $\omega=2\omega_{1}$. Further, note that in the denominator of eqn.(\ref{eqn:DiracRGF}), the intermediate configuration energy is negative and the states at Dirac points are fixed at zero energy, such that the fluctuation scale $\omega_{1}$ for the renormalized lower band states is negative: $\omega_{1}<0$. Thus, in order to put the RG of the spinon dispersion on the same footing as the RG relations that led to the phase diagram Fig.(\ref{fig:phasediagram}), we will write the RG equation (\ref{eqn:DiracRGF}) in terms of the fluctuation scale $\omega$ used earlier
\begin{eqnarray}
\epsilon_{1\Lambda}^{n-1}-\epsilon_{1\Lambda}^n =\frac{V_0^2}{2( |\epsilon_{1\Lambda_n}| +\frac{V_0-|\omega|}{2})}~.
\label{eqn:DiracRGFF}
\end{eqnarray}
As the dispersion $\epsilon_{1\Lambda}<0$, the RG flow in eqn\eqref{eqn:DiracRGFF} is RG relevant if $|\omega |-V_0>2|\epsilon_{1\Lambda_{n}}|$ and irrelevant for the converse. Recall that 
$\text{max}(|\epsilon_{1\Lambda_n}|)=1.0$, as can be seen from Fig.(\ref{fig:dispersion}) (by noting that the Dirac points will appear energy E=-1.0). Further, in order to stay within the boundaries of the gapless region $III$ in the phase diagram Fig. \ref{fig:phasediagram}, the maximum value of $|\omega |-V_0$ is $\text{max}(|\omega |-V_0)=\Delta_0 \sim 1.355$. From this, we can see that $|\omega |-V_0 < 2|\epsilon_{1\Lambda_{n}}|$~. This leads to the RG flow being irrelevant, and to a reduction of the dispersion magnitude until it stops at a stable fixed point value.
\par 
At the stable fixed point  
\begin{equation}
|\epsilon_{1\Lambda^{*}}|= (|\omega|-V_0)/2=\hbar v^{*}_{F} \Lambda^{*}~,
\end{equation}
giving the final renormalized Fermi velocity as $v^{*}_{F}=(|\omega|-V_0)/2\hbar\Lambda^{*}$~. In the continuum limit, the difference RG Eqn.(\ref{eqn:DiracRGFF}) is replaced by a differential RG flow equation
\begin{eqnarray}
\frac{d|\epsilon_{1\Lambda}|}{d \ln(\frac{\Lambda}{\Lambda_0})} &=& \frac{V_0^2}{2( |\epsilon_{1\Lambda_n}| +\frac{V_0-|\omega|}{2})}~,  
\label{eqn:RGDiracdisp}
\end{eqnarray}
where $d|\epsilon_{1\Lambda}|=|\epsilon_{1\Lambda}^{n-1}|-|\epsilon_{1\Lambda}^n|$ and $\ln(\frac{\Lambda}{\Lambda_0})$ is a negative quantity.
Integrating this differential equation between the bare $\Lambda_{0}$ to the final stable fixed point value $\Lambda^{*}$ obtained in eqn. (\ref{eqn:RGDiracdisp}), we have
\begin{eqnarray}
\frac{\Lambda^{*}}{\Lambda_0} &=& e^{\frac{1}{2V_0^2}(|\epsilon_{1\Lambda^{*}}|-|\epsilon_{1\Lambda_0}|)(V_0-|\omega|+2|\epsilon_{1\Lambda_0}|)}~.
\end{eqnarray}
In the above, $(|\epsilon_{1\Lambda^{*}}|-|\epsilon_{1\Lambda_0}|)$ is negative, such that for $|\epsilon_{1\Lambda_0}|=1.0$, $|\omega|=2\Delta_0$ and $V_0=\Delta_0$, the expression $(V_0-|\omega|+2|\epsilon_{1\Lambda_0}|)$ is positive and the overall exponent negative, leading to $\Lambda^{*}<\Lambda_{0}$.
\par
From the above, we can now derive the renormalisation of the Fermi velocity $v_{F}$. In the bare dispersion at a scale $\Lambda^{*}$ from the Dirac points, the energy expression is $\epsilon^{0}_{1\Lambda^{*}}=v^{0}_{F}\Lambda^{*}$, where $v^{0}_{F}$ is the bare Fermi velocity. From the above discussion, we find that corrections to the dispersion are RG irrelevant, such that $\epsilon_{1\Lambda^{*}}<\epsilon^{0}_{1\Lambda^{*}}$~, i.e., the overall magnitude of the dispersion at the stable fixed point is lower in comparison to the bare one. This leads us to conclude that the renormalized Fermi velocity is lowered from its bare value, $v^{*}_{F}=(|\omega|-V_0)/2\hbar\Lambda^{*} < v^{0}_{F}$, and the Dirac cones are flattened. Below, we will see the effects of velocity renormalisation on the quasiparticle residue as well as the lifetime of the spinon excitations. This will offer further insight on the nature of the gapless spinon liquid phase.
\subsection{Spinon self-energy, residue and lifetime}
From the stable fixed point single-spinon energy, we compute the real part of the self energy for the gapless phase from the renormalisation of the spinon dispersion (i.e., the difference between final fixed point energy and initial energy)
\begin{eqnarray}
\Sigma^{Re}_\Lambda(\omega) &=& \hbar v_{F}^{*}\Lambda-\hbar v_{F}^0\Lambda \nonumber \\
&=& \frac{(\omega-V_0)\Lambda}{2\hbar\Lambda^{*}}-v_f^0\Lambda\quad  \nonumber \\
&=& \frac{\Lambda(\omega-V_0)}{2\hbar\Lambda_0}\exp[\frac{1}{V_0^2}(\frac{\omega-V_0}{2}-1)^2]-v_f^0\Lambda~,
\end{eqnarray}
where we have used the expressions for $v_{F}^{*}$, $\Lambda^{*}$ and $|\epsilon_{1\Lambda^{*}}|$ computed above.
\par 
For $(\frac{\omega-V_0}{2}-1)<< V_{0}$, the exponent can be simplified to
\begin{eqnarray}
\Sigma^{Re}_\Lambda(\omega) 
&=& \frac{\Lambda}{2\Lambda_0}(\omega-C)~,
\end{eqnarray}
where $C=V_0+2v_f^0\Lambda_0$. As $\Delta_{0}\leq\omega\leq 2\Delta_{0}$~, the self energy can be written as
\begin{eqnarray}
\Sigma^{Re}_\Lambda(\omega)&=& \frac{\Lambda}{2\Lambda_0}(\omega-C)~\Theta(2\Delta_0-\omega)\Theta(\omega-\Delta_0) 
\end{eqnarray}
We can now compute the quasi-particle residue $Z(\Lambda,\omega)$ for low-energy excitations $\omega \to \Delta_{0}+$
\begin{eqnarray}
Z^{-1} = 1-\frac{d\Sigma^{Re} }{d\omega} = (1-\frac{\Lambda}{2\Lambda_0})~.
\end{eqnarray}
The quasiparticle residue $Z\to 1$ upon approaching the gapless Dirac-point Fermi surface ($\Lambda\to 0$), indicating non-interacting spinon quasiparticles in the gapless phase. However, in order to learn whether this result indicates a Landau-Fermi liquid or not, we should also compute the spinon lifetime $\tau$. Thus, from the Kramers-Kronig relation, we compute the imaginary part of the self-energy and thereby the spinon lifetime $\tau(\Lambda,\omega)$
\begin{eqnarray}
\Sigma^{Im}_\Lambda(\omega)&=&-\frac{1}{\pi}\mathcal{P}\int_{-\infty}^{\infty} \frac{\Sigma^{Re}_\Lambda(\omega')}{\omega'-\omega}~d\omega'\nonumber \\
&=& -\frac{1}{\pi}\int_{\Delta_0}^{2\Delta_0}\frac{ \frac{\Lambda}{2\Lambda_0}(\omega'-C) }  {\omega'-\omega}~d\omega'\nonumber \\
&=& -\frac{1}{\pi}\frac{\Lambda}{2\Lambda_0}\Big[\Delta_0 + (\omega-C)\ln \Big(\Big|\frac{2\Delta_0-\omega}{\Delta_0-\omega}\Big|\Big)\Big] \nonumber \\
&=& \frac{1}{\tau (\Lambda,\omega)}~(\text{as}\lim _{\Lambda\to 0}Z = 1)~, 
\end{eqnarray}
where $\mathcal{P}$ is the principal part of the contour integral. Then, upon taking the limit $\omega \to \Delta_{0}+$ in the gapless phase (and away from the quantum critical lines 2 and 3 in Fig.(\ref{fig:phasediagram})) 
\begin{eqnarray}
\lim_{\omega \to \Delta_{0}+}\lim_{\Lambda\to 0} \tau (\Lambda,\omega)^{-1} \propto \lim_{\omega \to \Delta_{0}+}\lim_{\Lambda\to 0} \frac{\Lambda}{\Lambda_{0}}\to 0~,
\end{eqnarray}
indicating that the lifetime $\tau$ diverges upon approaching the Dirac-point Fermi surface, without matching the faster ($1/\Lambda^{2}$) divergence of the Landau-Fermi liquid. This departure from the standard Fermi liquid paradigm arises from an interplay of the inter-spinon interactions and the vanishing spinon density of states as the Dirac point Fermi surface is approached.
\par
On the other hand, upon taking the limits in the other way around, we can obtain the lifetime for the excitations on the quantum critical lines 2 and 3
\begin{eqnarray}
\lim_{\Lambda\to 0}\lim_{\omega \to \Delta_{0}+} \frac{1}{\tau (\Lambda,\omega)} \propto 
\frac{\Lambda}{\Lambda_{0}}(C-\Delta_{0})\log(\Delta_{0}-\omega)\sim  \infty~,
\end{eqnarray}
indicating that the spinons have a vanishing lifetime precisely on the quantum critical line. The spinon Greens function in the gapless phase can thus be written as
\begin{eqnarray}
G (\Lambda,\omega) = \frac{1}{(\Delta_0-\omega)+v_f^0\Lambda-\Sigma^{Re}-i\Sigma^{Im}}~.
\end{eqnarray}
The Fourier transformation of $G(\Lambda,\omega)$ shows that the two-point correlation function decays algebraically in real space, supporting the formation of an {\it algebraic spin liquid} ground state with likely long-ranged entanglement in phase III. 
\section{Discussion and Outlook}
The RG phase diagram shown in Fig.(\ref{fig:phasediagram}) encapsulates the major findings of this work. We conclude with a discussion on the relevance of our findings, as well as an outlook on future prospects. Experiments on the Herbertsmithite and Volborthite materials are thought to be described by the physics of the $S=1/2$ Heisenberg KA. Indeed, recent experiments on Volborthite at high magnetic fields ($\sim 20-160T$) have revealed the existence of a robust plateau in the magnetisation per site of $1/3$~\cite{PhysRevB.98.020404}. From our RG analysis, such a gapped ground state for the Heisenberg KA corresponds to a quantum spin liquid driven by spinon correlations (phase-II in Fig.(\ref{fig:phasediagram})). It will, therefore, be interesting to search for experimental signatures of topological order within the plateau, e.g., fractional excitations above the gap. Indeed, claims of the observation of such fractional excitations in Herbetsmithite at zero-field from neutron scattering measurements have appeared recently~\cite{han2012fractionalized}. A magnetisation plateau at $1/3$ in the Heisenberg KA at finite field has also been proposed from numerical methods~\cite{nishimoto2013controlling,PhysRevB.88.144416}. 
\par
Equally interesting would be a search for the gapless algebraic spin liquid found by us (phase-III in Fig.(\ref{fig:phasediagram})). For this, it may be possible to melt the gapped spin liquid in phase-II by enhancing quantum fluctuations through the application of pressure~\cite{PhysRevB.84.174424,PhysRevB.96.075139}. We note that the existence of a gapless algebraic phase in the KA has just recently been proposed as being accessible via thermal melting of the finite-field ordered gapped phase~\cite{chen2017finite}. It will be interesting to compare such a thermally induced gapless phase with that reached purely from critical quantum fluctuations (line-2 in Fig.(\ref{fig:phasediagram})). Theoretical proposals of a gapless Dirac spin liquid in the Heisenberg KA at zero field have appeared in, for instance, the slave-fermion mean-field approach of Hermele \emph{et al.}\cite{PhysRevB.77.224413} as well as in large-scale DMRG simulations of He \emph{et al.}\cite{PhysRevX.7.031020}. It is pertinent to check whether the gapless phase observed by us at finite field is connected in any way to these results. Further, it is promising to investigate whether such a 2D algebraic Dirac spin liquid can exists at the boundaries of a topologically ordered ground state of the three dimensional pyrochlore $S=1/2$ Heisenberg antiferromagnet~\cite{PhysRevB.83.205101,vafek2014dirac}. 
\par
The non-trivial multicritical point lying at the intersection of the three spin liquid phases (black circle in Fig.(\ref{fig:phasediagram})) represents an intriguing result. The experimental observation of such an exotic quantum critical point would be highly valuable, as it is likely to be masked by the formation of a symmetry-broken ground state. Finally, given that critical theories describing line-2 and line-3 relate to relativistic fermionic criticality in two dimensions, it appears pertinent to employ similar RG methods to the investigation of Mott criticality in graphene and its analogs~\cite{gonzalez1994non,PhysRevB.59.R2474,PhysRevB.78.205433,PhysRevB.80.075432}), as well as topological transitions in quantum Hall systems~\cite{hatsugai1997topological} and other topological insulators~\cite{mohanta2014emergent}~.
\section{Acknowledgement}
The authors acknowledge S. Patra, A. Panigrahi, S. Bandopadhyay, G. Dev Mukherjee, V. Ravi Chandra, D. Sen, S. Rao and R. K. Singh for several enlightening discussions. S. Pal and A. Mukherjee acknowledge CSIR, Govt. of India and IISER Kolkata for financial support. S. L. thanks the DST, Govt. of India for funding through a Ramanujan Fellowship during which this project was initiated.
\appendix
\section{Scattering processes in PP and PH channels}
\label{appendix-I}
\begin{figure}[h!]
\centering
\includegraphics[scale=.13]{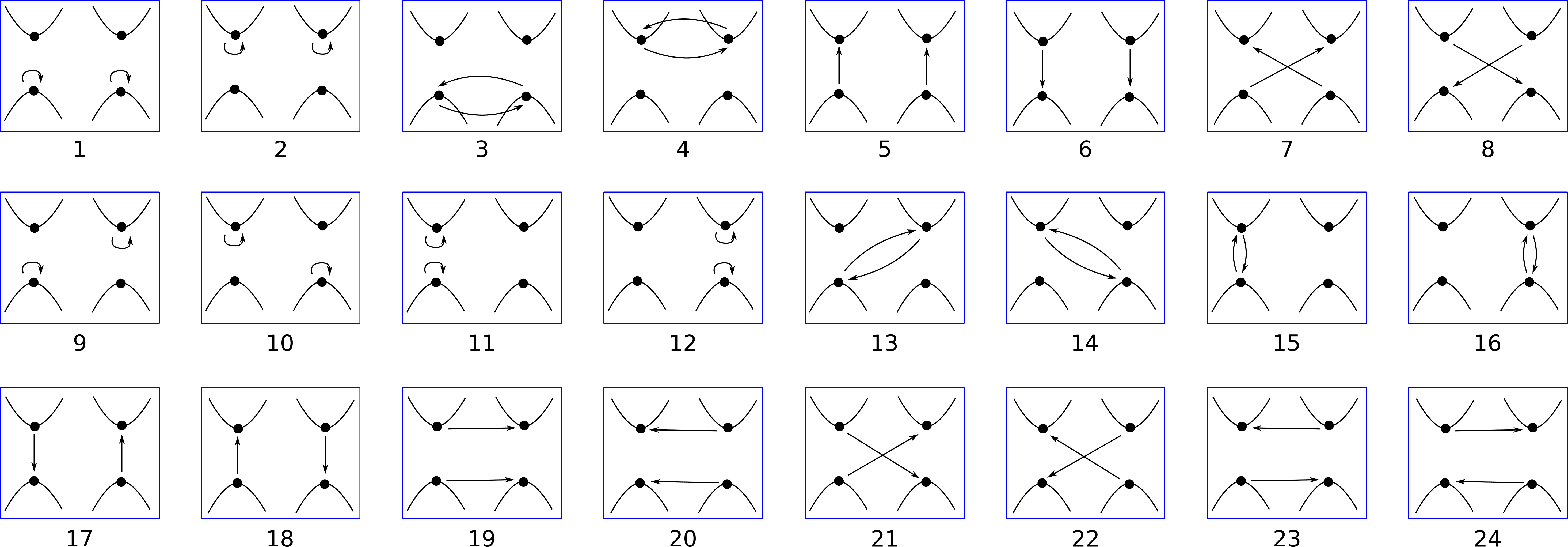}
\caption{Schematic diagram for various scattering processes between the patch-centres (black circles) of the dispersion spectrum. Processes 1, 2, 5 and 6 involve zero momentum-transfer, whereas processes 3, 4, 7 and 8 involve a non-zero momentum-transfer in the PP scattering channel. Similarly, diagrams 9-24 represent the different scattering processes in the PH channel.}
\label{Fig:HDspinonterm}
\end{figure}

\noindent
The processes labelled as (1-8) and (9-24) in Fig.(\ref{Fig:HDspinonterm}) present all possible intra- and inter-band scattering responsible for the instability in the PP for PH channels respectively.
\par 
The pseudo-spinors for the PP channel are define as
\begin{eqnarray}
f^+_{(1/2)D}=[C^{\dagger}_{k(1/2)b}~ C_{k(1/2)a}],~ f^-_{(1/2)D}=\begin{bmatrix}
C_{k(1/2)b}\\ C^{\dagger}_{k(1/2)a}
\end{bmatrix}
\end{eqnarray} 
and the corresponding pseudospins are
\begin{eqnarray}
A^z_{(1/2)D} &=& f^+_{(1/2)D} \frac{\sigma^z}{2}f^-_{(1/2)D}=\frac{1}{2}(\hat{n}_{k(1/2)b}+\hat{n}_{k(1/2)a}-1)\nonumber \\
A^+_{(1/2)D} &=& f^+_{(1/2)D} \frac{\sigma^+}{2}f^-_{(1/2)D}=C^{\dagger}_{k(1/2)b} C^{\dagger}_{k(1/2)a}\nonumber \\
A^-_{(1/2)D} &=& f^+_{(1/2)D} \frac{\sigma^-}{2}f^-_{(1/2)D}=C_{k(1/2)a} C_{k(1/2)b}~.
\end{eqnarray}
Similarly, the pseudo-spinors for PH channel are defined as
\begin{eqnarray}
f^+_{(1/2)S}=[C^{\dagger}_{k(1/2)b}~ C^\dagger_{k(1/2)a}],~ f^-_{(1/2)S}=\begin{bmatrix}
C_{k(1/2)b}\\ C_{k(1/2)a}
\end{bmatrix}
\end{eqnarray} 
and the corresponding pseudo-spins are
\begin{eqnarray}
A^z_{(1/2)S} &=& f^+_{(1/2)S} \frac{\sigma^z}{2}f^-_{(1/2)S}=\frac{1}{2}(\hat{n}_{k(1/2)b}-\hat{n}_{k(1/2)a})\nonumber \\
A^+_{(1/2)S} &=& f^+_{(1/2)S} \frac{\sigma^+}{2}f^-_{(1/2)S}=C^{\dagger}_{k(1/2)b} C_{k(1/2)a}\nonumber \\
A^-_{(1/2)S} &=& f^+_{(1/2)S} \frac{\sigma^-}{2}f^-_{(1/2)S}=C^\dagger_{k(1/2)a} C_{k(1/2)b}~.
\end{eqnarray}
Thus, the pseudo-spins $\vec{A}_{(1/2)D}$ and $\vec{A}_{(1/2)S}$ are spin-1/2 objects in the PP ($\hat{n}_{\vec{k}_{1a}}=\hat{n}_{\vec{k}_{1b}}$) and PH ($\hat{n}_{\vec{k}_{1a}}+\hat{n}_{\vec{k}_{1b}}=1$) subspaces respectively. The index $(1/2)$ of the pseudo-spinors (or pseudospins) the represents the (lower/upper) band. One can check that the components of the pseudospins ($\vec{A}_{D/S}$) satisfy the $SU(2)$ spin algebra. 
%and we will see that the interaction part of the total  Hamiltonian is also SU(2) symmetric. 
Using  these pseudospins in the Hamiltonian for the scattering in the PP channel, eqn.(\ref{eqn:HD}), we obtain
\begin{eqnarray}
H_D &=& \epsilon_1(2A^z_{1D} +1) + \epsilon_2(2A^z_{2D} +1)
+ V_{q=0}\Big(A^+_{1D}A^-_{1D} +(1\leftrightarrow 2)\Big)
- V_{q\neq 0}\Big(A^+_{1D}A^-_{1D} -(1\leftrightarrow 2)\Big)\nonumber \\
&&+ V_{q=0}\Big(A^+_{2D}A^-_{1D} +(1\leftrightarrow 2)\Big)
- V_{q\neq 0}\Big(A^+_{2D}A^-_{1D} -(1\leftrightarrow 2)\Big)\nonumber \\
&=& (\epsilon_1+\epsilon_2+V)(A^z_{1D}+A^z_{2D})+(\epsilon_2-\epsilon_1)(A^z_{2D}-A^z_{1D})+ V(A^+_{2D}A^-_{1D} + \text{h.c.})~,
\label{Eqn:holondoublon}
\end{eqnarray}
where, in the last step, we have used $A^+_{D}A^-_{D}=1/2+A^z_{D}$. As discussed in the main text, $V(=V_{q=0}-V_{q\neq 0})>0$ as $V_{q\neq 0}<0$. Similarly, we can use $\hat{n}_{k(1/2)a}=A_{(1/2)D}^z-A_{(1/2)S}^z+\frac{1}{2}$,  $\hat{n}_{k(1/2)b}=A_{(1/2)D}^z+A_{(1/2)S}^z+\frac{1}{2}$ in the Hamiltonian for scattering in the PH channel, eqn.(\ref{eqn:spinon}), to obtain (upto a constant)
\begin{eqnarray}
H_S=2V A^z_{2D}A^z_{1D}-2V \vec{A}_{2S}\vec{A}_{1S}+V(A^z_{2D}+A^z_{1D})~.
\end{eqnarray}
Then, the total Hamiltonian for the system is ((eqn.(\ref{twopatchpseudospin}) in the main text)
\begin{eqnarray}
H &=& H_D+H_S \nonumber\\
&=& (\epsilon_1+\epsilon_2+2V)(A^z_{2D}+A^z_{1D})+ (\epsilon_2-\epsilon_1)(A^z_{2D}-A^z_{1D})+ 2V(\vec{A}_{2D}\vec{A}_{1D}-\vec{A}_{2S}\vec{A}_{1S})\nonumber\\
&=& U(A^z_{2D}+A^z_{1D})+ \Delta (A^z_{2D}-A^z_{1D})
+2V(\vec{A}_{2D}\vec{A}_{1D}-\vec{A}_{2S}\vec{A}_{1S})~,
\label{Eqn:FinalH}
\end{eqnarray}
where $U=\epsilon_1+\epsilon_2+2V$ and $\Delta=\epsilon_2-\epsilon_1 $.
%==================================================================================
\section{Two particle interaction RG}
\label{appendix-II}
\noindent
Inter-band scattering processes between the patch-centres and the boundaries lead to renormalization of the two particle interaction strength in the two-patch Hamiltonian. The scattering involves both excitation as well as de-excitation processes. In an excitation process, a pseudospin scatters between the lower band patch-centre and the boundary of the upper band, giving
\begin{eqnarray}
&&\frac{V_n^2 A^+_{2D,2h} A^-_{1D,1\Lambda}|\uparrow_{1\Lambda}\uparrow_{2\Lambda}\rangle\langle \uparrow_{1\Lambda}\uparrow_{2\Lambda}|A^+_{2D,2\Lambda} A^-_{1D,1h}}{\langle \uparrow_{1\Lambda}\uparrow_{2\Lambda}|(\omega - H_{\Lambda})|\uparrow_{1\Lambda}\uparrow_{2\Lambda}\rangle}\nonumber \\
&&=\frac{V_n^2 A^+_{2D,2h}A^-_{1D,1h}|\downarrow_{1\Lambda}\uparrow_{2\Lambda}\rangle\langle \uparrow_{1\Lambda}\downarrow_{2\Lambda}|}{\omega-(2V_0+\frac{V_n}{2})}~.
\label{eqn:RGI2}
\end{eqnarray}
On the other hand, in a de-excitation process, a pseudospin scatters between the boundary of the lower band and patch-centre of the upper band, yielding
\begin{eqnarray}
&&\frac{V_n^2 A^+_{1D,1h} A^-_{2D,2\Lambda}|\uparrow_{1\Lambda}\uparrow_{2\Lambda}\rangle\langle \uparrow_{1\Lambda}\uparrow_{2\Lambda}|A^+_{1D,1\Lambda} A^-_{2D,2h}}{\langle \uparrow_{1\Lambda}\uparrow_{2\Lambda}|(\omega - H_{\Lambda})|\uparrow_{1\Lambda}\uparrow_{2\Lambda}\rangle}\nonumber \\
&&=\frac{V_n^2 A^+_{1D,1h}A^-_{2D,2h}|\uparrow_{1\Lambda}\downarrow_{2\Lambda}\rangle\langle \downarrow_{1\Lambda}\uparrow_{2\Lambda}|}{\omega-(2V_0+\frac{V_n}{2})}~.
\label{eqn:RGI1}
\end{eqnarray}
Then, the total contribution from both types of processes gives
\begin{widetext}
\begin{eqnarray}
&&\frac{V_n^2}{\omega-(2V_0+\frac{V_n}{2})}(A^+_{1D,1h}A^-_{2D,2h}|\uparrow_{1\Lambda}\downarrow_{2\Lambda}\rangle\langle \downarrow_{1\Lambda}\uparrow_{2\Lambda}| 
+ A^+_{2D,2h}A^-_{1D,1h}|\downarrow_{1\Lambda}\uparrow_{2\Lambda}\rangle\langle \uparrow_{1\Lambda}\downarrow_{2\Lambda}|)\nonumber \\
&=& \frac{K}{2}[(A^+_{1D,1h}A^-_{2D,2h}+A^+_{2D,2h}A^-_{1D,1h})(\uparrow_{1\Lambda}\downarrow_{2\Lambda}\rangle\langle \downarrow_{1\Lambda}\uparrow_{2\Lambda}|+|\downarrow_{1\Lambda}\uparrow_{2\Lambda}\rangle\langle \uparrow_{1\Lambda}\downarrow_{2\Lambda}|)\nonumber \\
&&+ (A^+_{1D,1h}A^-_{2D,2h}-A^+_{2D,2h}A^-_{1D,1h})(\uparrow_{1\Lambda}\downarrow_{2\Lambda}\rangle\langle \downarrow_{1\Lambda}\uparrow_{2\Lambda}|
 -|\downarrow_{1\Lambda}\uparrow_{2\Lambda}\rangle\langle \uparrow_{1\Lambda}\downarrow_{2\Lambda}|)]~,
\label{eqn:RGinteraction}
\end{eqnarray}
\end{widetext}
where $K=\frac{V_n^2}{\omega-(2V_0+\frac{V_n}{2})}$~. If we contract eqn.(\ref{eqn:RGinteraction}) with the state $\frac{1}{\sqrt{2}}(|\uparrow_{1\Lambda}\downarrow_{2\Lambda}\rangle + |\downarrow_{1\Lambda}\uparrow_{2\Lambda}\rangle$, we obtain
\begin{eqnarray}
&&\frac{K}{2}(A^+_{1D,1h}A^-_{2D,2h}+A^+_{2D,2h}A^-_{1D,1h})
+\frac{K}{2}(A^+_{1D,1h}A^-_{2D,2h}-A^+_{2D,2h}A^-_{1d,1h})~,
\end{eqnarray}
and the corresponding RG equation
\begin{eqnarray}
\delta V_n=\frac{2~(V_n/2)^2}{\omega-(2V_0+\frac{V_n}{2})}~.
\label{eqn:RGV}
\end{eqnarray}
For phase-I of the phase diagram (Fig.(\ref{fig:phasediagram})), we have numerically integrated the RG relation 
$\frac{dV}{d ln\Lambda}=\frac{d\Delta}{d ln\Lambda}$ with the limits of integration being $V_{0}\leq V \leq V^{*}$ and $\Delta_{0}\leq \Delta \leq \Delta^{*}$, where $V^{*}=2(\omega-2V_0)$ is the interaction at the the one-spinon gapped fixed point, and $\Delta^{*}$ is the fixed point hybridization gap to be determined. In this way, we find $\Delta^{*} = (\Delta_0+2\omega-5V_0)$, and plot it in terms of the (dark-to-light) pink colour bar in Fig.(\ref{fig:phasediagram}).
Similarly, phase-II is obtained by using the corresponding final fixed point value of the PH channel interaction (a measure of the strength of the two-spinon gap)~, $V^{S\ast}= 2(\Delta_0-\omega)$~.
%=========================================================================================
\section{Topological quantum numbers}
\label{appendix-III}
\noindent
We can use the spinon Greens function $\tilde{G}_{0}$ in the complex plane to define the topological quantum number $N_{\Lambda}$~\cite{phillips2012advanced}
\begin{eqnarray}
N_\Lambda &=& \frac{1}{2\pi i}\int dz~ \tilde{G}_0(z)\nonumber \\
&=& \frac{1}{2\pi i}\int dz~ \tilde{G}_0(z)~\partial_z \tilde{G}^{-1}_0(z) \qquad (\because \partial_z \tilde{G}^{-1}_0(z)=1) \nonumber \\
&=& \frac{1}{2\pi i}\int_{\tilde{C}} dz~ \frac{\partial}{\partial_z } ~\ln\Bigg(\frac{\tilde{G}^{-1}_0(z)}{\tilde{G}^{\ast -1}_0(z)}\Bigg)\nonumber \\
&=& \frac{1}{2\pi i}\Big[\ln\Bigg(\frac{\tilde{G}^{-1}_0(\infty)}{\tilde{G}^{\ast -1}_0(\infty)}\Bigg)-\ln\Bigg(\frac{\tilde{G}^{-1}_0(0)}{\tilde{G}^{\ast -1}_0(0)}\Bigg)\Big]~,
\end{eqnarray}
where $\tilde{C}$ is the upper half-circle in the complex plane. In eqn.(\ref{eqn:complexG}), for  $ \Delta_0> 2V_0$ and $z\rightarrow \infty, \quad \tilde{G}^{-1}_0(\infty) $ is always positive (i.e. $e^{i2\pi}$) and $\tilde{G}^{-1}_0(0)$ always negative (i.e. $e^{i\pi}$). On the other hand, for $ \Delta_0< 2V_0$ and $z\rightarrow \infty$, both $\tilde{G}^{-1}_0(\infty)$ and $\tilde{G}^{-1}_0(0)$ are positive (i.e. $e^{i2\pi}$), leading to
\begin{eqnarray}
N_\Lambda &=& \frac{1}{2\pi i} (4\pi i - 2\pi i)=1 \qquad \text{if} \quad \Delta_0> 2V_0~,\nonumber \\
 &=& \frac{1}{2\pi i} (4\pi i - 4\pi i)=0 \qquad \text{if} \quad \Delta_0< 2V_0~.
\end{eqnarray}
\par 
In order to compute the Chern number related to the one-spinon gap, we can write an effective two-level Hamiltonian in the low-energy neighbourhood of the gapped two-band problem as follows
\begin{eqnarray}
H=\begin{bmatrix}
\Delta_0/2 & v_F(p-p_F)\\
v_F(p-p_F) & -\Delta_0/2
\end{bmatrix}
=\vec{p}\cdot\vec{\sigma}~,
\end{eqnarray}
where the $\sigma$'s are Pauli matrices and $p$ is the momentum (with respect to the Fermi momentum). It is well known that the two-level problem (with a structure of the Hamiltonian given above) possesses a geometric (Berry) phase $\gamma=\Omega/2$, where $\Omega$ is the solid angle created by the loop on the Bloch sphere. If we take an integral over the entire Bloch sphere (with a total solid angle of $4\pi$), we obtain the topological Chern number (C)~\cite{RevModPhys.82.1959}
\begin{eqnarray}
C=\frac{1}{4\pi} 4\pi=1~.
 \end{eqnarray}
%=========================================================================================
\section{Twist Operators}
\label{appendix-IV}
\noindent
In section \ref{hybgaprenorm}, we have defined the twist operator projected onto the PP subspace (see eqns.(\ref{eqn:twist}) and (\ref{eqn:twistscattering})). The action of the projected doubled twist operator, $P_{D}\hat{O}^{2}P_{D}$, on a single particle state gives
\begin{eqnarray}
P_{D}\hat{O}^{2}P_{D}|k\rangle &=& P_{D}\hat{O}^{2}P_{D} \frac{1}{\sqrt{Vol}}\sum_{\vec{r}}e^{i\vec{k}\cdot\vec{r}}|1_{\vec{r}}\rangle \nonumber\\
&=&\frac{1}{\sqrt{Vol}}\sum_{\vec{r}}e^{i\vec{k}\cdot\vec{r}}P_{D}\hat{O}^2C^\dagger_{\vec{r}}\hat{O}^{\dagger 2}\hat{O}^2 P_{D}|0_{\vec{r}}\rangle \nonumber\\
&=& \frac{1}{\sqrt{Vol}}\sum_{\vec{r}}e^{i(\vec{k}-\frac{2\pi}{N_1})\vec{r}}|1_{\vec{r}}\rangle= |k-\frac{2\pi}{N_1}\rangle
\label{eqn:doubletwist}
\end{eqnarray}
In the last step we have used the result: $P_{D}\hat{O}^2C^\dagger_{\vec{r}}\hat{O}^{\dagger 2}P_{D}=e^{-i\frac{4\pi}{N_1}\vec{r}}C^\dagger_{\vec{r}}$ and $P_{D}\hat{O}^2 P_{D}|0_{\vec{r}}\rangle=e^{i\frac{2\pi}{N_1}\vec{r}}|0_{\vec{r}}\rangle$. As defined in the main text, $P_{D}=(16/9)\vec{A}^{2}_{1D}\vec{A}^{2}_{2D}$ is the projection operator on the PP channel for the 3rd and the 4th bands. In this way, the effect of doubled twist operator on the single spinon state ($|0_{\vec{r}}\rangle$) is to shift its momentum by $2\pi/N_1$. Defining $\vec{P}_{cm}$ as the center of mass momentum,
\begin{widetext}
\begin{eqnarray}
P_{D}\hat{O}^2\vec{P}_{cm}\hat{O}^{\dagger 2}P_{D}&=& P_{D}\hat{O}^2\sum_{\vec{k}_1}^{\vec{k}_2-2\pi/N_1}\vec{k}(\hat{n}_{\vec{k},1}+\hat{n}_{\vec{k},2})\hat{O}^{\dagger 2}P_{D}\nonumber\\
&=& P_{D}\hat{O}^2\vec{k}_1(n_{\vec{k}_1,1}+n_{\vec{k}_1,2})+(\vec{k}_1+\frac{2\pi}{N_1})(n_{\vec{k}_1+2\pi/N_1, 1}+n_{\vec{k}_1+2\pi/N_1, 2})+...\nonumber\\ &&+(\vec{k}_2-\frac{2\pi}{N_1})(n_{\vec{k}_2-2\pi/N_1, 1}+n_{\vec{k}_2-2\pi/N_1, 2})\hat{O}^{\dagger 2}P_{D}\nonumber\\
&=&\vec{k}_1(n_{\vec{k}_1+2\pi/N_1, 1}+n_{\vec{k}_1+2\pi/N_1, 2})+(\vec{k}_1+\frac{2\pi}{N_1})(n_{\vec{k}_1+4\pi/N_1, 1}+n_{\vec{k}_1+4\pi/N_1, 2})+...\nonumber\\
&&+(\vec{k}_2-\frac{4\pi}{N_1})(n_{\vec{k}_2-2\pi/N_1, 1}+n_{\vec{k}_2-2\pi/N_1, 2})+(\vec{k}_2-\frac{2\pi}{N_1})(n_{\vec{k}_2, 1}+n_{\vec{k}_2,2})\nonumber\\
&=&  \vec{k}_1(n_{\vec{k}_1+2\pi/N_1, 1}+n_{\vec{k}_1+2\pi/N_1, 2})+(\vec{k}_1+\frac{2\pi}{N_1})(n_{\vec{k}_1+4\pi/N_1, 1}+n_{\vec{k}_1+4\pi/N_1, 2})+...\nonumber\\
&&+(\vec{k}_2-\frac{4\pi}{N_1})(n_{\vec{k}_2-2\pi/N_1, 1}+n_{\vec{k}_2-2\pi/N_1, 2})+(\vec{k}_2-\frac{2\pi}{N_1})(n_{\vec{k}_1, 1}+n_{\vec{k}_1, 2})~,
\end{eqnarray}
\end{widetext}
where $\vec{k}_1$ and $\vec{k}_2$ are the momenta of the two Dirac-points. In the above, we used the fact that $\vec{k}_2\equiv \vec{k}_1$, i.e., these two points are connected by a reciprocal lattice vector of the magnetic Brillouin zone for the spinon system. The indices 1 and 2 represent, as always, the lower and upper bands respectively. Under the action of the doubled twist operator on the center of mass momentum, $\vec{k}_1\leftrightarrow \vec{k}_2-2\pi/N_1$, which is equivalent to the scattering of a spinon between the two patch-centres (i.e., the two Dirac-points). This scattering process is captured by the effective Hamiltonian
\begin{eqnarray}
H_{D} &=& V[A^+_{2D}A^-_{1D}+h.c.]\nonumber\\
&=& P_{D}(\hat{O}^2+\hat{O}^{\dagger 2})P_{D}~.
\end{eqnarray}
A similar set of arguments is employed in reaching the effective scattering Hamiltonian between the two patch-centres in the PH channel
\begin{eqnarray}
H_{S} &=& V[A^+_{2S}A^-_{1S}+h.c.]\nonumber\\
&=& P_{S}(\hat{O}^2+\hat{O}^{\dagger 2})P_{S}~,
\end{eqnarray}
where $P_{S}$ is the projection onto the same bands mentioned above, but for the PH channel. In the projected space of the 3rd and 4th bands the Hamiltonians $H_{D}$ and $H_{S}$ have a band-inversion symmetry, as the chemical potential is pinned at the middle of the two bands. 
%=========================================================================================
\section{RG relations in the gapless phase}
\label{appendix-V}
\noindent
We can write the RG equation for process-I (see Fig.(\ref{fig:DiracRG})) from
\begin{eqnarray}
&& \frac{V_0^2 A^+_{1D,1\Lambda_n}A^-_{2D,2\Lambda}A^+_{2D,2\Lambda}A^-_{1D,1\Lambda_n}}{\omega_1+\epsilon_{1\Lambda_n}+\frac{V_0}{2}}\nonumber \\
&=& \frac{V_0^2}{\omega_1+\epsilon_{1\Lambda_n}+\frac{V_0}{2}}(\frac{1}{2}+ A^z_{1D,1\Lambda_n})(\frac{1}{2}- A^z_{2D,2\Lambda})\nonumber \\ 
&=& \frac{V_0^2}{\omega_1+\epsilon_{1\Lambda_n}+\frac{V_0}{2}}(\frac{1}{2}- A^z_{2D,2\Lambda})~,
\end{eqnarray}
by taking the trace over the state $|1\Lambda_n\rangle$~. Similarly, the RG equation for process-II is obtained from
\begin{eqnarray}
&& \frac{V_0^2 A^-_{2D,2\Lambda_n}A^+_{1D,1\Lambda}A^-_{1D,1\Lambda}A^+_{2D,2\Lambda_n}}{\omega_2-\epsilon_{2\Lambda_n} -\frac{V_0}{2}}\nonumber \\
&=& \frac{V_0^2}{\omega_2-\epsilon_{2\Lambda_n} -\frac{V_0}{2}}(\frac{1}{2}+ A^z_{1D,1\Lambda})(\frac{1}{2}- A^z_{2D,2\Lambda_n})\nonumber \\
&=& \frac{V_0^2}{\omega_2-\epsilon_{2\Lambda_n} -\frac{V_0}{2}}(\frac{1}{2}+ A^z_{1D,1\Lambda})~,
\end{eqnarray}
by taking the trace over the state $|2\Lambda_n\rangle$~.
%=========================================================================================
\bibliography{bibliography}
\end{document}